\documentclass[aps,prl,superscriptaddress,twocolumn,letterpaper,10pt,amsfonts,amsmath,amssymb]{revtex4-2}

\usepackage{graphicx}
\usepackage{hyperref}
\usepackage{xcolor}
\usepackage{color}
\hypersetup{
    colorlinks,
    linkcolor={blue!80!black},
    citecolor={blue!80!black},
    urlcolor={blue!80!black}
}
\usepackage{physics}
\usepackage{braket}
\usepackage{comment}
\usepackage{placeins}
\usepackage{siunitx}
\usepackage{chemformula}
\usepackage{lipsum}
\usepackage{titletoc}

\begin{document}

\title{Finite‐momentum superconductivity from chiral bands in twisted MoTe$_2$}

\newcommand{\equalcontrib}{These authors contributed equally.}

\author{Yinqi Chen}\thanks{\equalcontrib}
\affiliation{Hearne Institute of Theoretical Physics, Department of Physics \& Astronomy, Louisiana State University, Baton Rouge LA 70803, USA}

\author{Cheng Xu}\thanks{\equalcontrib}
\affiliation{Department of Physics and Astronomy, University of Tennessee, Knoxville, TN 37996, USA}

\author{Yang Zhang}
\affiliation{Department of Physics and Astronomy, University of Tennessee, Knoxville, TN 37996, USA}
\affiliation{Min H. Kao Department of Electrical Engineering and Computer Science, University of Tennessee, Knoxville, TN 37996, USA}

\author{Constantin Schrade}
\affiliation{Hearne Institute of Theoretical Physics, Department of Physics \& Astronomy, Louisiana State University, Baton Rouge LA 70803, USA}

\date{\today}
\date{\today}

\begin{abstract}
A recent experiment has reported unconventional superconductivity in twisted bilayer MoTe$_2$, emerging from a normal state that exhibits a finite anomalous Hall effect -- a signature of intrinsic chirality. 
Motivated by this discovery, we construct a continuum model for twisted MoTe$_2$ constrained by lattice symmetries from first-principles calculations that captures the moir\'{e}-induced inversion symmetry breaking even in the absence of a displacement field. 
Building on this model, we show that repulsive interactions give rise to finite-momentum superconductivity via the Kohn-Luttinger mechanism in this chiral moir\'{e} system. 
Remarkably, the finite-momentum superconducting state can arise solely from internal symmetry breaking of the moir\'{e} superlattice, differentiating it from previously studied cases that require external fields.
 It further features a nonreciprocal quasiparticle dispersion and an intrinsic superconducting diode effect.  
Our results highlight a novel route to unconventional superconducting states in twisted transition metal dichalcogenides moir\'{e} systems, driven entirely by intrinsic symmetry-breaking effects.
\end{abstract}

\maketitle
Moir\'{e} systems based on transition metal dichalcogenides (TMDs) have in recent years emerged as a versatile platform for realizing a wide range of interaction-driven electronic phases~\cite{kennes2021moire,mak2022semiconductor}.
Examples include correlated insulators~\cite{PhysRevLett.121.026402,PhysRevB.102.201115,tang2020simulation,ghiotto2021quantum,li2021continuous,xu2022tunable}, generalized Wigner crystals~\cite{regan2020mott,li2021imaging,jin2021stripe}, quantum anomalous Hall states~\cite{li2021quantum}, fractional quantum anomalous Hall states~\cite{park2023observation,PhysRevX.13.031037}, and fractional quantum spin Hall states~\cite{kang2024evidence}. 
Remarkably, superconductivity has also recently been observed in twisted bilayer WSe$_2$ at twist angles $\theta = 3.56^\circ$~\cite{xia2025superconductivity} and $\theta = 5^\circ$~\cite{guo2025superconductivity}, motivating intense theoretical efforts to understand the underlying pairing mechanisms~\cite{PhysRevB.110.035143,PhysRevB.111.014507,PhysRevB.111.L060501,christos2024approximate,guerci2024topological,tuo2024theory,qin2024kohn,fischer2024theory,PhysRevResearch.4.043048,PhysRevResearch.5.L012034,PhysRevLett.134.136503,kim2025theory,zhu2025plane}.
In particular, at $\theta=5^\circ$, the emergence of superconductivity near a van Hove singularity and an antiferromagnetic metal has pointed to spin fluctuations as a possible mediator of pairing between time-reversed states in opposite valleys~\cite{PhysRevB.110.035143,PhysRevB.111.014507,PhysRevB.111.L060501,christos2024approximate,guerci2024topological,tuo2024theory,qin2024kohn,fischer2024theory}.

\begin{figure}[!t]
    \centering
    \includegraphics[width=\linewidth]{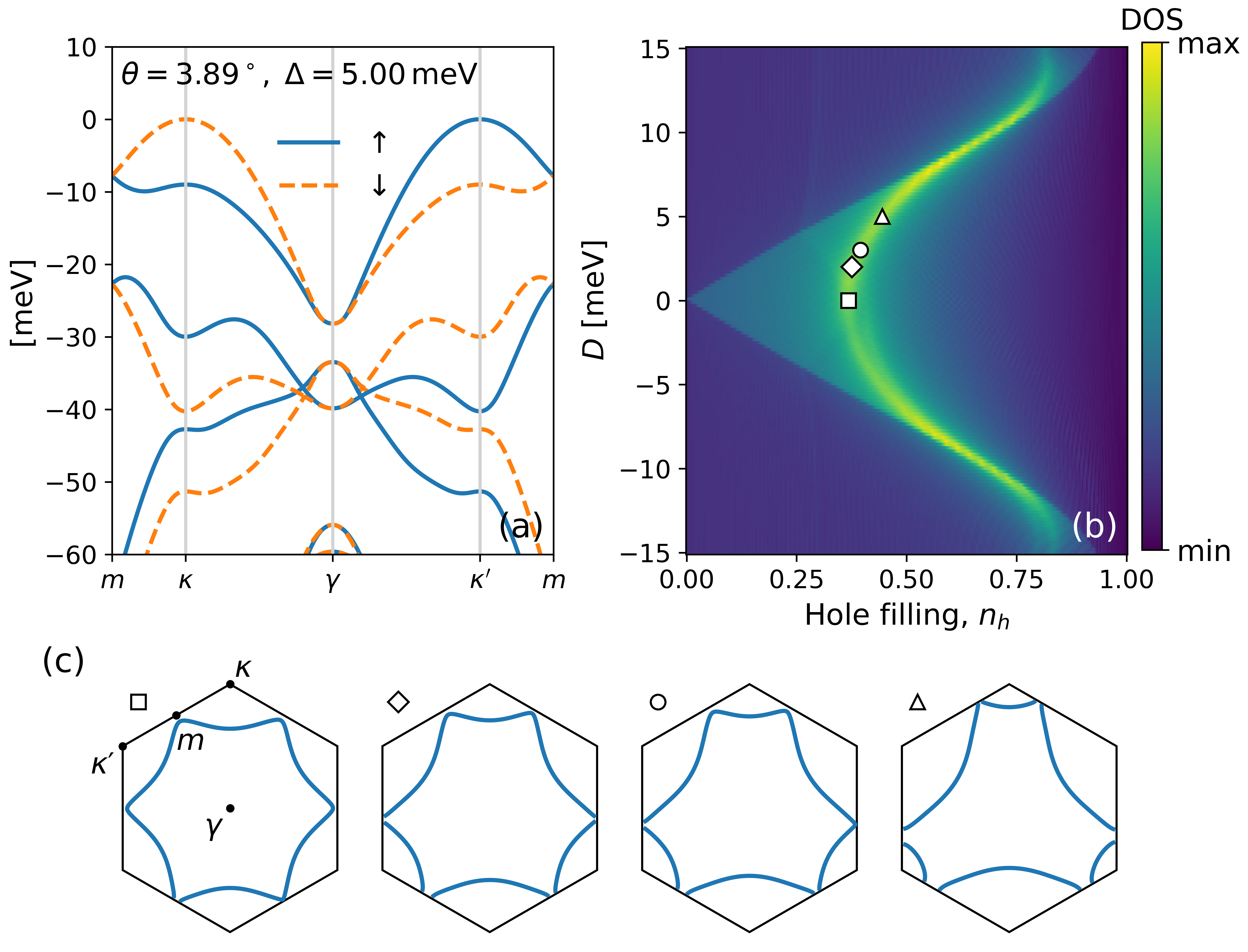}
    \caption{
(a) Band structure of tMoTe$_2$ along a path in the mBZ at $\theta = 3.89^\circ$. 
A displacement field $D = 5$\,meV induces a sizable spin splitting, with spins ($\uparrow,\downarrow$) locked to valleys ($\tau = \pm$).
(b) DOS of the topmost valence band, $\xi_{\boldsymbol{k},+}\equiv\xi_{\boldsymbol{k}}$, 
as a function of its hole density, $n_{h,+}\equiv n_{h}$ and $D$. 
(c) Fermi surfaces for representative values of $(D, n_h)$. From left to right: $D = \{0, 2, 3, 5$\}\, meV, $n_h = \{0.368, 0.376, 0.395, 0.445\}$.
    }
    \label{fig:1}
\end{figure}

Very recently, signatures of superconductivity have been observed for another member moir\'{e} TMD family, 
twisted bilayer MoTe$_2$ (tMoTe$_2$) at $\theta \sim 3.8^\circ$~\cite{xu2025signatures} and have attracted much interest~\cite{xu2025chiral,shi2025,nosov2025,huang2025,guerci2025,zhang2025,hu2025layer}.
Intriguingly, the state above the transition temperature to the zero-resistance state exhibits an anomalous Hall effect and magnetic hysteresis, signaling intrinsic chirality and time-reversal symmetry breaking. Moreover, this time-reversal symmetry breaking coexists with the breaking of inversion symmetry by the moir\'{e} pattern~\cite{zhang2024polarization} and an applied 
displacement field. From these findings raise many fundamental questions arise:
What microscopic mechanisms can give rise to superconductivity in a twisted TMD system from a parent normal state that is already chiral?  What is the nature of the possible superconducting order? And what experimental probes could reveal its symmetry-breaking properties? 

In this work, we address these questions by proposing and studying a Kohn-Luttinger mechanism for superconductivity in tMoTe$_2$. Our analysis is based on a new continuum model for the normal state of tMoTe$_2$ fitted from first-principles calculations, that captures the intrinsic inversion symmetry breaking of the moir\'{e} pattern even in the absence of an external displacement field. Based on this revised continuum model, we show that tMoTe$_2$ can host a finite-momentum superconducting state, which is distinct from previously studied cases and does not rely on externally applied symmetry-breaking fields. Moreover, we also demonstrate that the superconducting state exhibits a nonreciprocal quasiparticle spectrum, Bogoliubov Fermi surfaces, and an intrinsic superconducting diode effect.
We hope that our results with contribute to an understanding on the nature of superconductivity in chiral transition metal dichalcogenides moir\'{e} systems.

\textit{Normal-state.}
The band structure of tMoTe$_2$ is generally described by a continuum model 
that captures the low-energy bands near the $\pm K$ valleys~\cite{WuFengcheng2019,xu2024maximally,mao2024transfer,Yujin2024}. These valleys are separated by a large momentum transfer and can thus be treated as two decoupled degrees of freedom, labelled by an index $\tau = \pm$. 
The Hamiltonian near the $\tau$-valley takes the form of a $2 \times 2$ matrix in
the space of the two layers, $\ell=t,b$, 
\begin{equation}
H_{\tau} = \begin{pmatrix}
\varepsilon_{t,\tau}(\hat{\boldsymbol{k}}) + \Delta_+(\boldsymbol{r}) + \frac{D}{2} & \Delta_{T,\tau}(\boldsymbol{r}) \\
\Delta_{T,\tau}^\dagger(\boldsymbol{r}) & \varepsilon_{b,\tau}(\hat{\boldsymbol{k}}) + \Delta_-(\boldsymbol{r}) - \frac{D}{2}.
\end{pmatrix}    
\label{Eq1}
\end{equation}
Here, the diagonal entries describe the kinetic energy and intra-layer moir\'{e} potentials, while the off-diagonal entries describe inter-layer tunneling. The displacement field, $D$, enters as a potential between the layers.

We now describe the terms in Eq.\,\eqref{Eq1} in more detail: First, the kinetic terms, 
$
\varepsilon_{\ell,\tau}(\hat{\boldsymbol k})
=
-\hbar^{2}[\hat{\boldsymbol k}-\tau \boldsymbol K_{\ell}]^{2}/(2 m^{*})
$, model the lowest bands of the uncoupled layers near their respective Dirac points, $\boldsymbol K_{\ell}$. These bands are spin-polarized due to the strong Ising spin-orbit coupling and carry opposite spin-polarization for the two valleys as enforced by time-reversal symmetry. Here, we will adopt the choice $\boldsymbol{K}_t = ( -1/\sqrt{3}, 1/2 ) \cdot (4\pi / 3a_M)$ and $\boldsymbol{K}_b = (0,1) \cdot (4\pi / 3a_M)$ where $a_M \approx a / \theta$ is the moir\'{e} period, $a = 3.52\,\text{\AA}$ is the monolayer lattice constant, and $\theta = 3.89^\circ$.
\\

Second, the intra-layer moir\'{e} potentials and inter-layer tunnelings are given by, 
\begin{align}
\Delta_\pm(\boldsymbol{r}) &= 2V_1 \sum_{i=1,3,5} \!\cos\left( \boldsymbol{g}_i^{1} \!\cdot\! \boldsymbol{r} \!\pm\! \phi_1 \right) + 2V_2 \!\sum_{i=1,3,5} \cos\left( \boldsymbol{g}_i^{2} \!\cdot\! \boldsymbol{r} \right)
,\nonumber \\ 
\Delta_T(\boldsymbol{r}) &= w_1 \sum_{i=1}^{3} e^{-i \boldsymbol{q}_i^{1} \cdot \boldsymbol{r}} 
+ w_2 \sum_{i=1}^{3} e^{-i \boldsymbol{q}_i^{2} \cdot \boldsymbol{r}},
\end{align}
where $V_{1,2}$ and $\phi$ are the amplitudes and phase of the moir\'{e} potentials, which couple plane-wave states within the same layer via the ``first-star" vectors $\boldsymbol{g}_i^{1}=\boldsymbol{G}_i \equiv (4\pi / \sqrt{3}a_M) \cdot (\cos[(i-2)\pi/3], \sin[(i-2)\pi/3])$ and ``second-star" vectors $\boldsymbol{g}_i^{2}=\boldsymbol{G}_i + \boldsymbol{G}_{i+1}$. Moreover, $w_1$ and $w_2$ are inter-layer tunnelings that couple states in different layers via $(\boldsymbol{q}_1^{\,1},\boldsymbol{q}_2^{\,1},\boldsymbol{q}_3^{\,1})=(0,\boldsymbol{G}_2,\boldsymbol{G}_3)$ and $(\boldsymbol{q}_1^{\,2},\boldsymbol{q}_2^{\,2},\boldsymbol{q}_3^{\,2})=(\boldsymbol{G}_2+\boldsymbol{G}_3,\boldsymbol{G}_1,\boldsymbol{G}_4)$. 

It is important to note that the continuum model construction followed Ref.~\cite{WuFengcheng2019} preserves an \emph{artificial} emergent inversion symmetry due to the restriction of interlayer tunneling to real values. In contrast, the actual lattice structure of twisted MoTe$_2$ retains only a twofold rotational symmetry about the $y$-axis ($C_{2y}$), as evidenced by the small band splitting along the $\Gamma$--$M$ direction in the DFT spectrum~\cite{mao2024transfer,Yujin2024,zhang2024polarization}. To more accurately capture the intrinsic inversion symmetry breaking---an essential ingredient for finite-momentum pairing---we incorporate complex interlayer hopping into our model. This modification allows for a more faithful reproduction of the DFT band structure.

In what follows, we will adopt the representative parameters $(\phi_1, V_1, w_1, V_2, w_2) = (-81.0^\circ, 8.3~\text{meV}, -8.4 \, e^{-i2.4}~\text{meV}, 5~\text{meV}, 8 \, e^{-i1.69}\text{meV})$.

To illustrate the properties of the continuum model, we show its spectrum along a path in the moir\'{e} Brillouin zone (mBZ) in Fig.~\ref{fig:1}(a). The two topmost valence bands, $\xi_{\boldsymbol{k},\pm}$, are well-isolated from other bands. At $D = 0$, the bands exhibit a small splitting due to the complex phase of $w_{1,2}$, which breaks inversion symmetry even in the absence of an external field. Applying a finite displacement field $D \neq 0$ enhances this inversion breaking and thus further increases the band splitting. The corresponding density of states (DOS), shown in Fig.~\ref{fig:1}(b), features a prominent peak whose height is sensitive to both the hole filling and the value of $D$. Representative Fermi surfaces for selected values of $D$ that ``follow" the peak in the DOS are shown in Fig.~\ref{fig:1}(c).

\begin{figure}[!t]
    \centering
    \includegraphics[width=\linewidth]{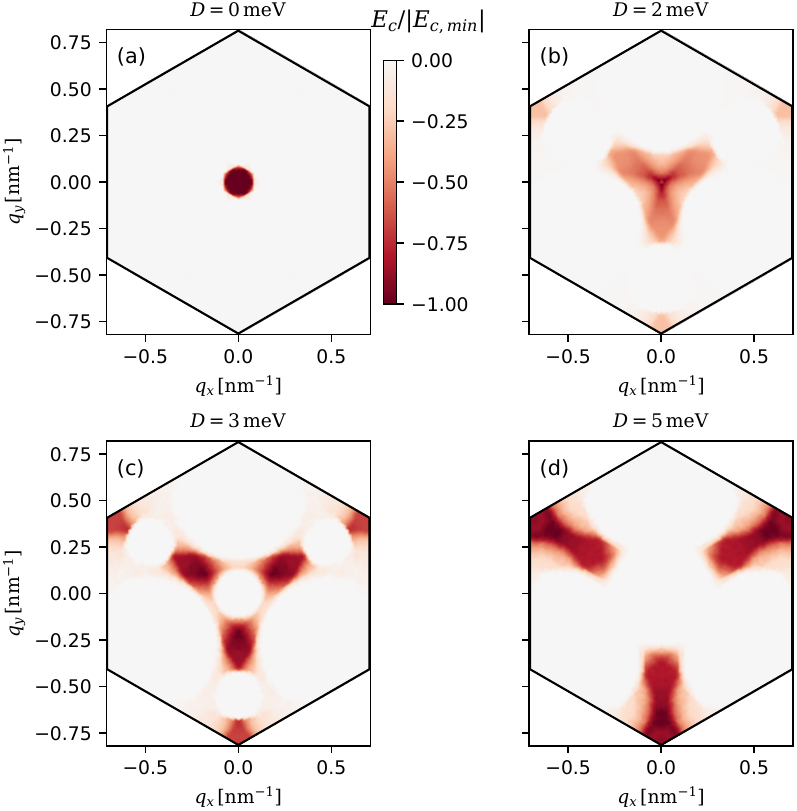}
    \caption{
(a) Condensation energy, $E_c = F - F_N$, versus Cooper-pair momentum $\boldsymbol{Q} = (Q_x, Q_y)$ at $n_h = 0.368$ and $D = 0$\,meV. The minimum slightly deviates from $\boldsymbol{Q} = \boldsymbol{\gamma}$ and appears at $\boldsymbol{Q} = (0.014,0.008)$ ${\rm nm}^{-1}$  and two $C_3$-symmetry-related momenta due to the weak inversion breaking from the complex phase in $w_{1,2}$.
(b-d) Same as (a), but for increasing $D$. The minima split into three $C_3$-related points, and an additional subleading minimum emerges near the $\boldsymbol{\kappa'}$-point. At $D=5$\,meV, the $\boldsymbol{\kappa'}$-point minimum becomes dominant.
    }
    \label{fig:2}
\end{figure}

\textit{Superconductivity.} Our first goal is to develop a microscopic theory of superconductivity in tMoTe$2$, assuming that time-reversal symmetry is spontaneously broken in the normal state such that only one valley band, $\xi_{\boldsymbol{k},+} \equiv \xi_{\boldsymbol{k}}$, remains ``active" and participates in pairing. In particular, we will propose and discuss a Kohn-Luttinger mechanism~\cite{kohn1965new,chubukov1993kohn}, where superconductivity emerges from overscreening of the Coulomb interaction.

We begin by defining the bare interaction Hamiltonian,
$
H_\mathrm{int}
=
(1/(2\Omega))
\sum_{\boldsymbol q}
V(\boldsymbol q)\hat\rho_{\boldsymbol q}\hat\rho_{-\boldsymbol q}
$. 
Here, $\Omega$ is the system area and 
$
V(\boldsymbol q)
= e^2\tanh(|\boldsymbol q|d)/(2\epsilon|\boldsymbol q|)
$
is the gate-screened Coulomb potential.
We choose $d = 10\,\text{nm}$ for the distance to the gates and $\epsilon = 5\epsilon_0$ for the dielectric permittivity ($\epsilon_0$ is the vacuum permittivity).
The bare interaction also includes the density operator projected on the $\xi_{\boldsymbol{k}}$-band, $
\hat{\rho}_{\boldsymbol{q}+\boldsymbol{G}} = \sum_{\boldsymbol{k}} F_{\boldsymbol{k}, \boldsymbol{k}+\boldsymbol{q}}(\boldsymbol{G})\, \hat{c}^\dagger_{\boldsymbol{k}} \hat{c}_{\boldsymbol{k}+\boldsymbol{q}}
$. 
Here, $\hat{c}_{\boldsymbol{k}}$ is an electron annihilation operator in this band 
and the form factors are 
$
F_{\boldsymbol{k}, \boldsymbol{k}+\boldsymbol{q}}(\boldsymbol{G}) = \sum_{\ell,\boldsymbol{G}'} u^*_{\ell,\boldsymbol{G}'}(\boldsymbol{k}) u_{\ell,\boldsymbol{G}'+\boldsymbol{G}}(\boldsymbol{k}+\boldsymbol{q})
$
with 
$u_{\ell,\boldsymbol{G}}(\boldsymbol{k})$ 
the Fourier coefficients of the periodic part of the Bloch wavefunctions, 
$
u_{\boldsymbol{k}}(\boldsymbol{r}) = \sum_{\boldsymbol{G}} u_{\ell,\boldsymbol{G}}(\boldsymbol{k}) e^{i\boldsymbol{G}\cdot\boldsymbol{r}}
|\ell\rangle
$.
\begin{figure}[!t]
    \centering
    \includegraphics[width=1.0\linewidth]{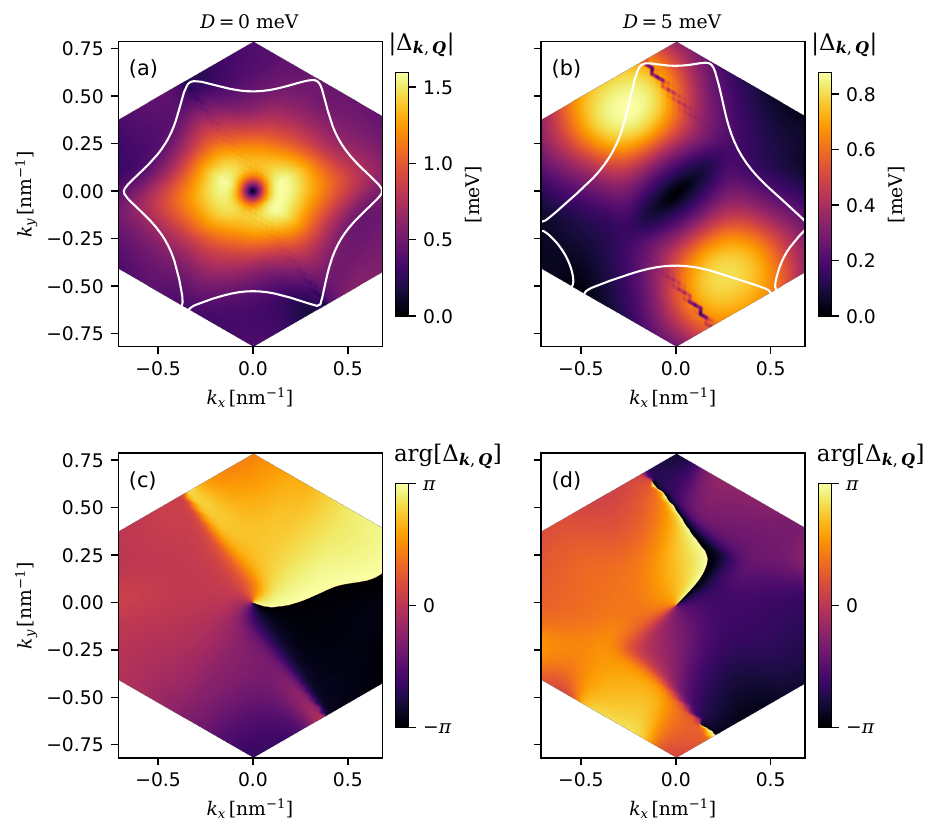}
    \caption{
(a) Magnitude of the superconducting order parameter $\Delta_{\boldsymbol{k},\boldsymbol{Q}}$ at $D = 0$\,meV and $n_h = 0.368$, evaluated at the Cooper-pair momentum $\boldsymbol{Q} = (0.014, 0.008)$ ${\rm nm}^{-1}$ that minimizes the condensation energy. The Fermi surface is shown as a white line.
(b) Same as (a) but for $D = 5$\,meV, $n_h=0.445$ and corresponding optimal $\boldsymbol{Q} = (0.567,0.328)$ ${\rm nm}^{-1}$.
(c) Phase of $\Delta_{\boldsymbol{k},\boldsymbol{Q}}$ at $D = 0$\,meV, showing a $(k_x + i k_y)$–like winding.
(d) Same as (c) but for $D = 5$\,meV.
    }
    \label{fig:3}
\end{figure}

As a next step, we incorporate the screening effects at the level of the random‐phase approximation (RPA). The resulting RPA‐dressed interaction is given by~\cite{xu2025chiral},
\begin{equation}
[V_{\text{RPA}}(\boldsymbol{q})]^{-1}_{\boldsymbol{G}\boldsymbol{G}'} = V^{-1}(\boldsymbol{q}+\boldsymbol{G})\delta_{\boldsymbol{G}\boldsymbol{G}'} - \Pi_{\boldsymbol{G}\boldsymbol{G}'}(\boldsymbol{q}), 
\label{Eq3}
\end{equation}
with
$\Pi_{\boldsymbol G,\boldsymbol G'}(\boldsymbol q)
= (1/\Omega)\sum_{\boldsymbol k}
F_{\boldsymbol k,\boldsymbol k+\boldsymbol q}(\boldsymbol G)
F^*_{\boldsymbol k,\boldsymbol k+\boldsymbol q}(\boldsymbol G')
[n_F(\xi_{\boldsymbol k})-n_F(\xi_{\boldsymbol k+\boldsymbol q})]/
(\xi_{\boldsymbol k}-\xi_{\boldsymbol k+\boldsymbol q}) 
$ being the static polarization and $n_F$ the Fermi-Dirac distribution.

For the superconducting instability, the relevant scattering processes involve two‐particle states, $\{|\boldsymbol{k}+\boldsymbol{Q}/2\rangle\otimes|\hspace{-2pt}-\boldsymbol{k}+\boldsymbol{Q}/2\rangle\}$, with finite center‐of‐mass momentum $\boldsymbol{Q}$. Accordingly, we introduce the effective Hamiltonian, 
\begin{equation}
\begin{split}
H_{\text{eff}} &= \sum_{\boldsymbol{k}}    
\xi_{\boldsymbol{k}}
\,
\hat{c}^\dagger_{\boldsymbol{k}}\hat{c}_{\boldsymbol{k}}
\\
&
+
\frac{1}{2\Omega} 
\sum_{\boldsymbol{k},\boldsymbol{k}',\boldsymbol{Q}}
g_{\boldsymbol{k},\boldsymbol{k}',\boldsymbol{Q}}\,
\hat c^{\dagger}_{\boldsymbol{k}+\frac{\boldsymbol{Q}}{2}}
\hat c^{\dagger}_{-\boldsymbol{k}+\frac{\boldsymbol{Q}}{2}}
\hat c_{-\boldsymbol{k}'+\frac{\boldsymbol{Q}}{2}}
\hat c_{\boldsymbol{k}'+\frac{\boldsymbol{Q}}{2}}
\end{split}
\label{Eq4}
\end{equation}
where we have defined the effective pairing kernel as,
$g_{\boldsymbol{k},\boldsymbol{k}',\boldsymbol{Q}} = \sum_{\boldsymbol{G},\boldsymbol{G}'} F_{-\boldsymbol{k}+\boldsymbol{Q}/2,-\boldsymbol{k}'+\boldsymbol{Q}/2}(\boldsymbol{G}) [V_{\text{RPA}}(\boldsymbol{k}-\boldsymbol{k}')]_{\boldsymbol{G},\boldsymbol{G}'} F_{\boldsymbol{k}+\boldsymbol{Q}/2,\boldsymbol{k}'+\boldsymbol{Q}/2}(-\boldsymbol{G}')$.

A mean-field decoupling of Eq.\,\eqref{Eq4} in the Cooper-channel now gives the self‐consistency equation for the superconducting order parameter, $\Delta_{\boldsymbol{k},\boldsymbol{Q}}$, 
\begin{equation}
\begin{split}
\Delta_{\boldsymbol{k},\boldsymbol{Q}}
=
&-
\frac{1}{\Omega}
\sum_{\boldsymbol{k}'}
g_{\boldsymbol{k},\boldsymbol{k}',\boldsymbol{Q}} \, 
\frac{
\Delta_{\boldsymbol{k}',\boldsymbol{Q}}
}{
2 \tilde{E}_{\boldsymbol{k}',\boldsymbol{Q}}
}
\\
\times
&
\frac{1}{2}
\left[
\tanh
\left(
\frac{\beta E_{\boldsymbol{k}',\boldsymbol{Q}}}{2}
\right)
+
\tanh
\left(
\frac{\beta E_{-\boldsymbol{k}',\boldsymbol{Q}}}{2}
\right)
\right].
\end{split}
\label{Eq5}
\end{equation}
where $
E_{\boldsymbol{k},\boldsymbol{Q}}
=
\xi_{\boldsymbol{k},\boldsymbol{Q},-}
+
\sqrt{
(\xi_{\boldsymbol{k},\boldsymbol{Q},+})^{2}
+
|
\Delta_{\boldsymbol{k},\boldsymbol{Q}}
|^{2}
}
$ is
the quasiparticle dispersion and $
\xi_{\boldsymbol{k},\boldsymbol{Q},\pm}
=
(
\xi_{\boldsymbol{k}+\frac{\boldsymbol{Q}}{2}}
\pm
\xi_{-\boldsymbol{k}+\frac{\boldsymbol{Q}}{2}}
)/2
$ 
are the symmetrized and antisymmetrized normal-state dispersions. In addition, we have introduced the reduced quasiparticle dispersion, 
 $\tilde{E}_{\boldsymbol{k}',\boldsymbol{Q}}=E_{\boldsymbol{k},\boldsymbol{Q}}-\xi_{\boldsymbol{k},\boldsymbol{Q},-}$.
 
\begin{figure}[!t]
    \centering
    \includegraphics[width=1.0\linewidth]{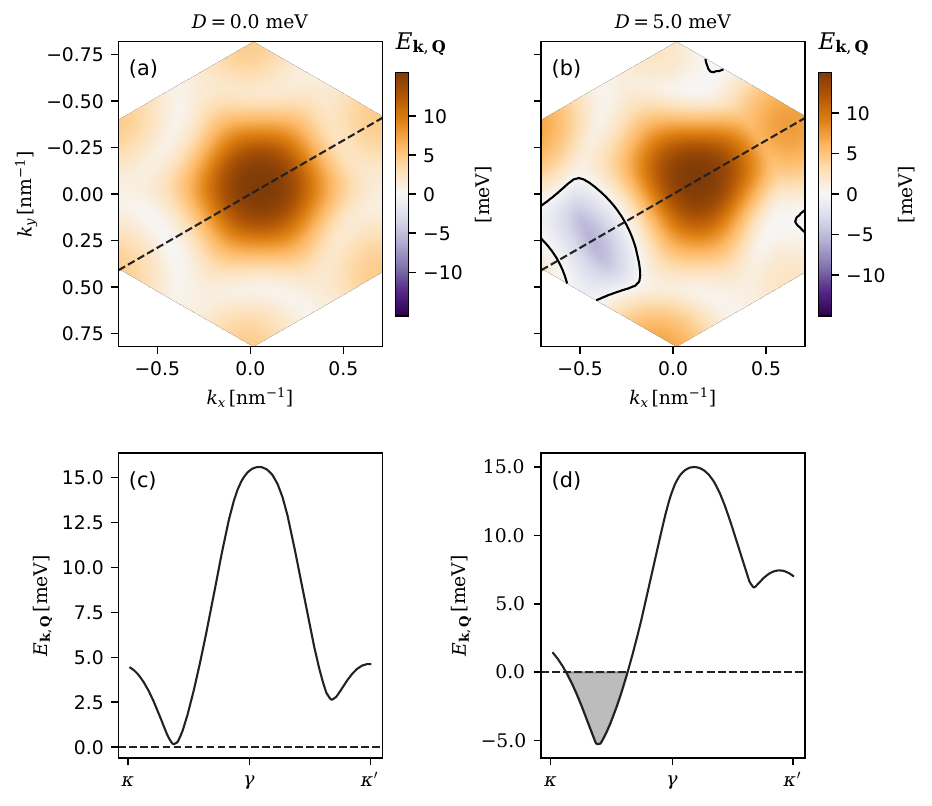}
    \caption{
(a) Quasiparticle dispersion $E_{\boldsymbol{k},\boldsymbol{Q}}$ at $D = 0$\,meV and optimal Cooper-pair momentum $\boldsymbol{Q} = (0.014, 0.008)$ ${\rm nm}^{-1}$.
(b) Same as (a), but for $D = 5$\,meV and corresponding optimal $\boldsymbol{Q} = (0.567, 0.328)$ ${\rm nm}^{-1}$. A Bogoliubov Fermi surface appears, highlighted in black.
(c) Dispersion along the dashed line in (a), showing an asymmetry between forward and backward directions.
(d) Dispersion along the dashed cut in (c), showing a sign change that leads to the formation of the Bogoliubov Fermi surface.
    }
    \label{fig:4}
\end{figure}

To determine the superconducting order parameter, we now solve 
Eq.\,\eqref{Eq5} self-consistently by imposing by fixed electron density 
$n_{e} = \sum_{\boldsymbol{k}} \langle c^\dagger_{\boldsymbol{k}}c_{\boldsymbol{k}} \rangle / \Omega$  and Cooper-pair momentum, $\boldsymbol{Q}$. The realized order parameter is then found by minimizing 
the condensation energy $E_{c} = F - F_{N}$ where $F$ is the free energy in the superconducting state, 
\begin{align}
\label{Eq7}
F 
&= 
-
\sum_{\boldsymbol{k}}
\left\{
\frac{1}{\beta}
\ln
(
1 + e^{-\beta E_{\boldsymbol{k},\boldsymbol{q}}}
)
-
\frac{1}{2}
(
\xi_{\boldsymbol{k},\boldsymbol{q},+}
-
\tilde{E}_{\boldsymbol{k},\boldsymbol{q}}
)
\right.
\\
&\qquad-
\left.
\frac{
\abs{\Delta_{\boldsymbol{k},\boldsymbol{q}}}^2
}{
8 \tilde{E}_{\boldsymbol{k},\boldsymbol{q}}
}
\left[
\tanh
\left(
\frac{\beta E_{\boldsymbol{k},\boldsymbol{q}}}{2}
\right)
+
\tanh
\left(
\frac{\beta E_{-\boldsymbol{k},\boldsymbol{q}}}{2}
\right)
\right]
\right\}
\nonumber
\end{align}
and $F_{N}$ is the free energy in the normal state. 

Our results are shown in Fig.~\ref{fig:2}. For $D=0$\,meV, the condensation energy is minimized near $(Q_{x},Q_{y}) = (0.014,0.008)$ ${\rm nm}^{-1}$, or at two equivalent momenta related by a $C_{3}$-rotation, see Fig.~\ref{fig:2}(a). This behavior is expected because the complex phases in $w_{1,2}$ weakly breaks the inversion symmetry of the normal state. As the displacement field $D$ increases, the minima split further but remain related by $C_{3}$-symmetry, as shown in Figs.~\ref{fig:2}(b) and (c). This additional splitting reflects the stronger inversion-symmetry breaking induced by $D$. Notably, another local minimum emerges near the $\kappa'$ point of the Brillouin zone. Although this minima is subleading for small $D$, it becomes the global minimum for $D \approx 5$\,meV, leading to a sizable Cooper-pair momentum near $\kappa'$, as depicted in Fig.~\ref{fig:2}(d).

To explain the origin of this transition, Fig.~\ref{fig:3} shows the superconducting order parameter and corresponding Fermi surfaces at $D=0$\,meV and $D=5$\,meV. At $D=0$\,meV, the Fermi surface is centered effectively around the $\gamma$ point, with comparable arc segments near $\kappa$ and $\kappa'$. As $D$ increases, the arcs near $\kappa$ contract while those near $\kappa'$ expand,
so that eventually pairing around $\kappa'$ can more effectively lower the system's condensation energy, as is seen in our numerical results.

\textit{Bogoliubov spectrum.} To further characterize the finite-momentum superconducting state, we now assume the system spontaneously selects one of the three $C_{3}$-related free-energy minima, $\boldsymbol{Q}$, and compute the Bogoliubov quasiparticle spectrum.

Two example plots of the Bogoliubov spectrum are shown in Fig.~\ref{fig:3}. Their properties can be understood from symmetry arguments: In the normal state, $\xi_{\boldsymbol{k}}$ is invariant under $C_{3}$ rotations and mirror reflections $\sigma_v$ along planes at angles $\Theta \in \{30^\circ,\,90^\circ,\,150^\circ\}$, where $\Theta=0^\circ$ is the positive $k_{x}$-axis. Once a superconducting order with a finite Cooper-pair momentum $\boldsymbol{Q}$ develops, this symmetry is reduced. Specifically, the quasiparticle spectrum is only symmetric along a fixed momentum direction, $E_{\boldsymbol{k},\boldsymbol{q}}=E_{-\boldsymbol{k},\boldsymbol{q}}$, if $\sigma_v\boldsymbol{k} = -\boldsymbol{k}$ and $\sigma_v\boldsymbol{q} = \boldsymbol{q}$.

This property can be directly seen from our numerical results. 
In Fig.~\ref{fig:3}, the Cooper-pair momentum $\boldsymbol{Q}$ lies along $\Theta=30^\circ$. Consequently, the direction $\Theta=120^\circ$ preserves a symmetric Bogoliubov dispersion, whereas all other directions, in particular $\Theta=30^\circ$, exhibit a marked asymmetry between positive and negative momenta. This asymmetry is a defining characteristic of nonreciprocal superconductivity \cite{davydovageier2024,chen2025,deazambuja2025}.

We can further refine this discussion by comparing the Bogoliubov spectrum at zero and finite displacement field, $D$. Most notably, we see that the aforementioned asymmetry of the Bogoliubov spectrum is present even if $D=0$, as can been from Fig.~\ref{fig:3}(a) and (c). Hence, no external displacement field is required in this case to realize non-reciprocal superconductivity. Instead, the nonreciprocal superconductivity arises directly from the intrinsic inversion and time-reversal symmetry breaking of the material. Moreover, as $D$ increases, the asymmetry in the Bogoliubov spectrum is enhanced, eventually giving rise to a Bogoliubov Fermi surface, as shown in Fig.~\ref{fig:3}(b) and (d).

\begin{figure}[!t]
    \centering
    \includegraphics[width=1.0\linewidth]{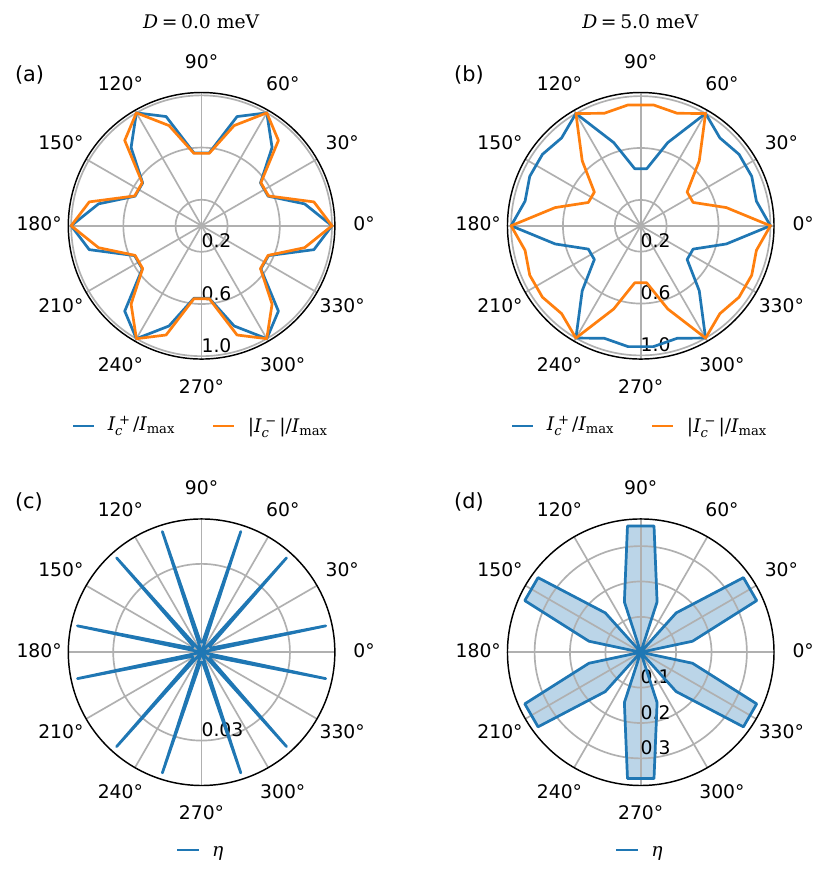}
    \caption{
(a) Angular dependence of the forward and backward critical currents, $I_c^+$ and $I_c^-$, at $D = 0$\,meV. The asymmetry arises from the inversion breaking due to the complex phase in $w_2$.
(b) Same as (a), but for $D = 5$\,meV.
(c) Angular dependence of the diode efficiency, $\eta$, at $D = 0\,$meV due to the critical current difference in (a).
(d) Same as (c), but for $D = 5$\,meV.
    }
    \label{fig:5}
\end{figure}

These features of the Bogoliubov spectrum have direct experimental implications. The asymmetries in the Bogoliubov dispersion can be probed through angle-resolved conductance measurements in a transparent normal metal–superconductor junction. Specifically, if the normal lead is oriented so that current flows along the direction of the Cooper-pair momentum, $\boldsymbol{Q}$, then the resulting current-voltage ($I\text{-}V$)
characteristic will be asymmetric with respect to $V\rightarrow-V$. In comparison, if current flows perpendicular to $\boldsymbol{Q}$, then the $I\text{-}V$ characteristic is symmetric under $V\rightarrow-V$. 
Moreover, the Bogoliubov Fermi surfaces lead to an enhancement of the density of states around zero energy, which can be similarly detected with conductance measurements.

\textit{Superconducting diode effect.} 
Lastly, an additional property of the finite-momentum superconducting state in tMoTe$_2$ is an intrinsic superconducting diode effect. By definition, such a superconducting diode effect effect manifests as an asymmetry between the critical current in the forward direction, $I^{+}_{c}(\Theta)$, and that in reverse, $I^{-}_{c}(\Theta)$. The degree of asymmetry is quantified by the diode efficiency,
$\eta = \frac{|I^{+}_{c}-I^{-}_{c}|}{I^{+}_{c}+I^{-}_{c}}$. 

Superconducting diodes have been studied in a broad range of symmetry-broken superconducting systems~\cite{nadeem2023superconducting,ando_observation_2020,he_phenomenological_2022,yuan_supercurrent_2022,zhang_general_2022,misaki_theory_2021,davydova_universal_2022,souto2024tuning,zazunov2009anomalous,brunetti2013anomalous,pal2022josephson,baumgartner_supercurrent_2022,legg2022superconducting,lotfizadeh2023superconducting,costa2023microscopic,maiani2023nonsinusoidal,hess2023josephson,hou2023ubiquitous,kononov2020one,souto2022josephson,gupta2023gate,ciaccia2023gate,valentini2023radio,greco2023josephson,legg2023parity,cuozzo2023microwave,PhysRevApplied.21.064029}, including symmetry-broken quantum materials where an intrinsic diode effect can arise without the 
applications of symmetry-breaking external fields~\cite{ZeroFieldDiode,scammell_theory_2022,wu_field-free_2022,zhang2024angle,PhysRevLett.132.046003,diez2023symmetry,hu2023josephson,chen2025}. The diode effect in tMoTe$_2$ falls into this latter category and, interestingly, requires neither an applied displacement field nor a magnetic field.

Our results are shown in Fig.~\ref{fig:5}. In particular, as depicted in 
Fig.~\ref{fig:5}(a) and (c), a finite difference between forward and reverse critical currents and thus a nonzero $\eta$ even at zero displacement field, $D=0$. The emergence of such a finite $\eta$ can be explained by the breaking of inversion symmetry due to the complex phases of $w_{1,2}$ even if $D=0$. 
Moreover, at $D=0$, we find that the diode efficiency exhibits relatively narrow peaks along certain $\Theta$-directions where it is enhanced, see Fig.~\ref{fig:5}(d). For $D\neq0$, these peaks turn into broader lobes with enhanced efficiency at $\Theta \in \{30^\circ,\,90^\circ,\,150^\circ\}$ and suppression at $\Theta \in \{0^\circ,\,60^\circ,\,120^\circ\}$. 

\textit{Conclusion.} In this work, we have developed a theory for superconducting emerging from chiral bands in  tMoTe$_2$ based on a Kohn-Luttinger mechanism. In particular, we have found that the resulting superconducting state exhibits a finite Cooper-pair momentum, leading to striking properties such as a nonreciprocal quasiparticle spectrum, Bogoliubov Fermi surfaces, and an intrinsic superconducting diode effect that necessitates neither an applied electric nor magnetic field. We hope that our results contribute to the understanding of symmetry-broken superconducting states in moiré TMD systems and, more broadly, to other platforms~\cite{luescher2025}, such as rhombohedral tetralayer graphene, where a superconducting state with a possible spin-polarized chiral superconducting was realized experimentally observed~\cite{han2024} and theoretically studied~\cite{geier2024chiral,chou2024intravalley,yang2024topological,qin2024chiral,parra2025band,dong2025,jahin2024enhanced,wang2024chiral,yoon2025quarter,maymann2025,christos2025,sedov2025,lesser2025}.

\textit{Note added.} While finalizing this manuscript, we became aware of an independent study~\cite{hu2025layer} investigating finite-momentum pairing under an applied gating field in tMoTe$_2$.

\begin{widetext}

\newpage

\begin{center}
\large{\bf Supplemental Material to `Finite‐momentum superconductivity from chiral bands in twisted MoTe$_2$' \\}
\end{center}
\begin{center}Yinqi Chen$^{1}$, Cheng Xu$^{2}$,  Yang Zhang$^{2,3}$, and Constantin Schrade$^{1}$
\\
{\it $^{1}$Hearne Institute of Theoretical Physics, Department of Physics \& Astronomy, Louisiana State University, Baton Rouge LA 70803, USA}
\\
{\it $^{2}$Department of Physics and Astronomy, University of Tennessee, Knoxville, TN 37996, USA}
\\
{\it $^{3}$Min H. Kao Department of Electrical Engineering and Computer Science,
University of Tennessee, Knoxville, Tennessee 37996, USA}
\end{center}
In the Supplemental Material, we provide more details on the density function theory (DFT) bands and continuum model fitting.

\section{DFT bands and continuum model fitting}
The large-scale plane-wave basis first principle calculations are carried out with Perdew-Burke-Ernzerhof (PBE) functionals using the Vienna Ab initio simulation package (VASP). We choose the projector augmented wave potentials, incorporating 6 electrons for each of the Mo and Te atoms. During the structural relaxation, we set the plane wave cutoff energy and the energy convergence criterion to 250 eV and $1 \times 10^{-6}$ eV, respectively. The structure is fully relaxed when the convergence threshold for the maximum force experienced by each atom is less than 10 meV/$\AA$. The space group of the relaxed structures is $P321$ (No. 150), whose point group is generated by a two-fold rotational symmetry along $y$ axis ($C_{2y}$), and three-fold rotational symmetry along $z$ axis ($C_{3z}$). In the crystal momentum space, the $C_{2y}$ symmetry only protects two-fold degeneracies at the invariant lines or points within the Brillouin Zone, as defined by the relation $C_{2y}\mathbf{k} \rightarrow \mathbf{k}$. Within this invariant domain, the Hamiltonian commutes with the symmetry operation, allowing it to be block-diagonalized into two distinct sectors, each characterized by unique eigenvalues $\pm \pi$. Due to the constraints imposed by the symmetry, a band represented by $e^{i\pi}$ is inherently degenerate with another band represented by $e^{-i\pi}$, forming a doubly-degenerate band structure. Consequently, the only lines that encapsulate the $C_{2y}$ symmetries within the two-dimensional Brillouin zone are the $\Gamma K$ lines (satisfying $2\mathbf{k}_1 + \mathbf{k}_2 = 0$). When considering the $C_{3z}$ rotational symmetry, the lines that meet the conditions $\mathbf{k}_1 + 2\mathbf{k}_2 = 0$ and $\mathbf{k}_1 - \mathbf{k}_2 = 0$ also emerge as the symmetry-invariant lines. As a result, bands along the $\Gamma K$ and $M K$ lines are always doubly degenerate, while a clear splitting is observed along the $\Gamma M$ line as shown in Fig.~\ref{fig:dftbands}. To accurately capture this splitting, we introduce complex intralayer hopping in the continuum model. During the parameter fitting process, we focus on the topmost band, as shown below, and the first moiré band is reproduced with high fidelity.
\begin{figure}[!b]
   \centering
   \includegraphics[width=0.47\linewidth]{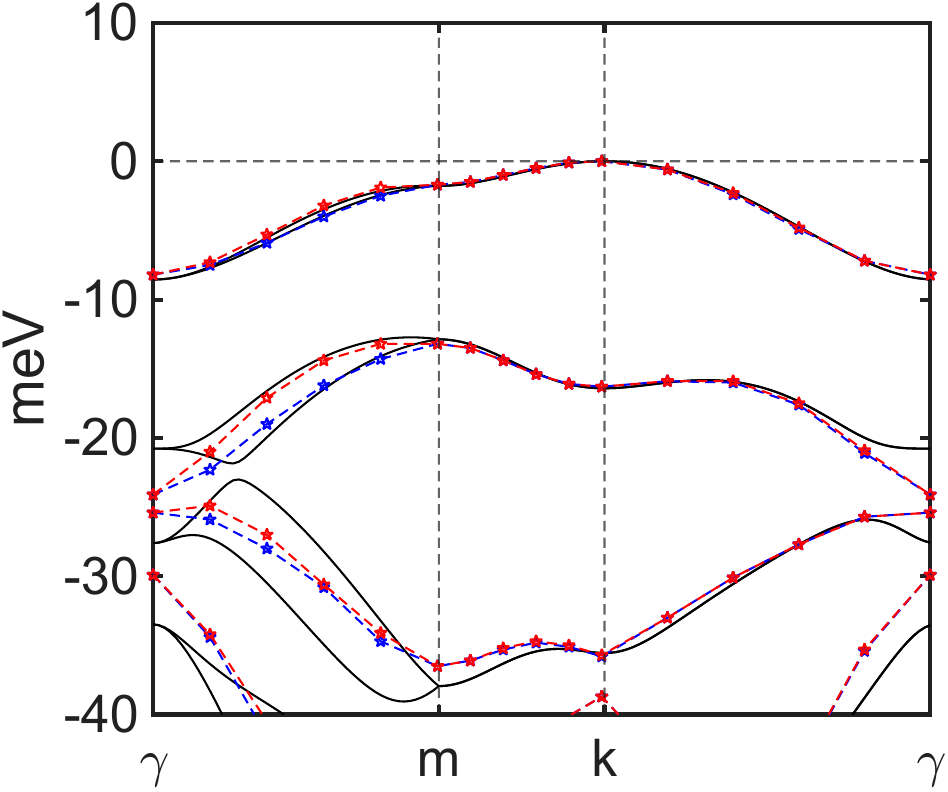}
   \caption{The black line represents the band structure calculated from the continuum model, while the dashed line corresponds to the DFT spectrum. Blue stars indicate spin-up states and red stars indicate spin-down states. All bands shown are for tMoTe$_2$ at $\theta = 3.15^\circ$.}
   \label{fig:dftbands}
\end{figure}

\end{widetext}


\begin{thebibliography}{97}
\expandafter\ifx\csname natexlab\endcsname\relax\def\natexlab#1{#1}\fi
\expandafter\ifx\csname bibnamefont\endcsname\relax
  \def\bibnamefont#1{#1}\fi
\expandafter\ifx\csname bibfnamefont\endcsname\relax
  \def\bibfnamefont#1{#1}\fi
\expandafter\ifx\csname citenamefont\endcsname\relax
  \def\citenamefont#1{#1}\fi
\expandafter\ifx\csname url\endcsname\relax
  \def\url#1{\texttt{#1}}\fi
\expandafter\ifx\csname urlprefix\endcsname\relax\def\urlprefix{URL }\fi
\providecommand{\bibinfo}[2]{#2}
\providecommand{\eprint}[2][]{\url{#2}}

\bibitem[{\citenamefont{Kennes et~al.}(2021)\citenamefont{Kennes, Claassen, Xian, Georges, Millis, Hone, Dean, Basov, Pasupathy, and Rubio}}]{kennes2021moire}
\bibinfo{author}{\bibfnamefont{D.~M.} \bibnamefont{Kennes}}, \bibinfo{author}{\bibfnamefont{M.}~\bibnamefont{Claassen}}, \bibinfo{author}{\bibfnamefont{L.}~\bibnamefont{Xian}}, \bibinfo{author}{\bibfnamefont{A.}~\bibnamefont{Georges}}, \bibinfo{author}{\bibfnamefont{A.~J.} \bibnamefont{Millis}}, \bibinfo{author}{\bibfnamefont{J.}~\bibnamefont{Hone}}, \bibinfo{author}{\bibfnamefont{C.~R.} \bibnamefont{Dean}}, \bibinfo{author}{\bibfnamefont{D.}~\bibnamefont{Basov}}, \bibinfo{author}{\bibfnamefont{A.~N.} \bibnamefont{Pasupathy}}, \bibnamefont{and} \bibinfo{author}{\bibfnamefont{A.}~\bibnamefont{Rubio}}, \href{https://www.nature.com/articles/s41567-020-01154-3}{\bibinfo{title}{Moir{\'e} heterostructures as a condensed-matter quantum simulator}}, \bibinfo{journal}{Nature Physics} \textbf{\bibinfo{volume}{17}}, \bibinfo{pages}{155} (\bibinfo{year}{2021}).

\bibitem[{\citenamefont{Mak and Shan}(2022)}]{mak2022semiconductor}
\bibinfo{author}{\bibfnamefont{K.~F.} \bibnamefont{Mak}} \bibnamefont{and} \bibinfo{author}{\bibfnamefont{J.}~\bibnamefont{Shan}}, \href{https://www.nature.com/articles/s41565-022-01165-6}{\bibinfo{title}{Semiconductor moir{\'e} materials}}, \bibinfo{journal}{Nature Nanotechnology} \textbf{\bibinfo{volume}{17}}, \bibinfo{pages}{686} (\bibinfo{year}{2022}).

\bibitem[{\citenamefont{Wu et~al.}(2018)\citenamefont{Wu, Lovorn, Tutuc, and MacDonald}}]{PhysRevLett.121.026402}
\bibinfo{author}{\bibfnamefont{F.}~\bibnamefont{Wu}}, \bibinfo{author}{\bibfnamefont{T.}~\bibnamefont{Lovorn}}, \bibinfo{author}{\bibfnamefont{E.}~\bibnamefont{Tutuc}}, \bibnamefont{and} \bibinfo{author}{\bibfnamefont{A.~H.} \bibnamefont{MacDonald}}, \href{https://link.aps.org/doi/10.1103/PhysRevLett.121.026402}{\bibinfo{title}{Hubbard model physics in transition metal dichalcogenide moir\'e bands}}, \bibinfo{journal}{Phys. Rev. Lett.} \textbf{\bibinfo{volume}{121}}, \bibinfo{pages}{026402} (\bibinfo{year}{2018}).

\bibitem[{\citenamefont{Zhang et~al.}(2020)\citenamefont{Zhang, Yuan, and Fu}}]{PhysRevB.102.201115}
\bibinfo{author}{\bibfnamefont{Y.}~\bibnamefont{Zhang}}, \bibinfo{author}{\bibfnamefont{N.~F.~Q.} \bibnamefont{Yuan}}, \bibnamefont{and} \bibinfo{author}{\bibfnamefont{L.}~\bibnamefont{Fu}}, \href{https://link.aps.org/doi/10.1103/PhysRevB.102.201115}{\bibinfo{title}{Moir\'e quantum chemistry: Charge transfer in transition metal dichalcogenide superlattices}}, \bibinfo{journal}{Phys. Rev. B} \textbf{\bibinfo{volume}{102}}, \bibinfo{pages}{201115} (\bibinfo{year}{2020}).

\bibitem[{\citenamefont{Tang et~al.}(2020)\citenamefont{Tang, Li, Li, Xu, Liu, Barmak, Watanabe, Taniguchi, MacDonald, Shan et~al.}}]{tang2020simulation}
\bibinfo{author}{\bibfnamefont{Y.}~\bibnamefont{Tang}}, \bibinfo{author}{\bibfnamefont{L.}~\bibnamefont{Li}}, \bibinfo{author}{\bibfnamefont{T.}~\bibnamefont{Li}}, \bibinfo{author}{\bibfnamefont{Y.}~\bibnamefont{Xu}}, \bibinfo{author}{\bibfnamefont{S.}~\bibnamefont{Liu}}, \bibinfo{author}{\bibfnamefont{K.}~\bibnamefont{Barmak}}, \bibinfo{author}{\bibfnamefont{K.}~\bibnamefont{Watanabe}}, \bibinfo{author}{\bibfnamefont{T.}~\bibnamefont{Taniguchi}}, \bibinfo{author}{\bibfnamefont{A.~H.} \bibnamefont{MacDonald}}, \bibinfo{author}{\bibfnamefont{J.}~\bibnamefont{Shan}}, \bibnamefont{et~al.}, \href{https://www.nature.com/articles/s41586-020-2085-3}{\bibinfo{title}{Simulation of hubbard model physics in wse$_2$/ws$_2$ moir{\'e} superlattices}}, \bibinfo{journal}{Nature} \textbf{\bibinfo{volume}{579}}, \bibinfo{pages}{353} (\bibinfo{year}{2020}).

\bibitem[{\citenamefont{Ghiotto et~al.}(2021)\citenamefont{Ghiotto, Shih, Pereira, Rhodes, Kim, Zang, Millis, Watanabe, Taniguchi, Hone et~al.}}]{ghiotto2021quantum}
\bibinfo{author}{\bibfnamefont{A.}~\bibnamefont{Ghiotto}}, \bibinfo{author}{\bibfnamefont{E.-M.} \bibnamefont{Shih}}, \bibinfo{author}{\bibfnamefont{G.~S.} \bibnamefont{Pereira}}, \bibinfo{author}{\bibfnamefont{D.~A.} \bibnamefont{Rhodes}}, \bibinfo{author}{\bibfnamefont{B.}~\bibnamefont{Kim}}, \bibinfo{author}{\bibfnamefont{J.}~\bibnamefont{Zang}}, \bibinfo{author}{\bibfnamefont{A.~J.} \bibnamefont{Millis}}, \bibinfo{author}{\bibfnamefont{K.}~\bibnamefont{Watanabe}}, \bibinfo{author}{\bibfnamefont{T.}~\bibnamefont{Taniguchi}}, \bibinfo{author}{\bibfnamefont{J.~C.} \bibnamefont{Hone}}, \bibnamefont{et~al.}, \href{https://www.nature.com/articles/s41586-021-03815-6}{\bibinfo{title}{Quantum criticality in twisted transition metal dichalcogenides}}, \bibinfo{journal}{Nature} \textbf{\bibinfo{volume}{597}}, \bibinfo{pages}{345} (\bibinfo{year}{2021}).

\bibitem[{\citenamefont{Li et~al.}(2021{\natexlab{a}})\citenamefont{Li, Jiang, Li, Zhang, Kang, Zhu, Watanabe, Taniguchi, Chowdhury, Fu et~al.}}]{li2021continuous}
\bibinfo{author}{\bibfnamefont{T.}~\bibnamefont{Li}}, \bibinfo{author}{\bibfnamefont{S.}~\bibnamefont{Jiang}}, \bibinfo{author}{\bibfnamefont{L.}~\bibnamefont{Li}}, \bibinfo{author}{\bibfnamefont{Y.}~\bibnamefont{Zhang}}, \bibinfo{author}{\bibfnamefont{K.}~\bibnamefont{Kang}}, \bibinfo{author}{\bibfnamefont{J.}~\bibnamefont{Zhu}}, \bibinfo{author}{\bibfnamefont{K.}~\bibnamefont{Watanabe}}, \bibinfo{author}{\bibfnamefont{T.}~\bibnamefont{Taniguchi}}, \bibinfo{author}{\bibfnamefont{D.}~\bibnamefont{Chowdhury}}, \bibinfo{author}{\bibfnamefont{L.}~\bibnamefont{Fu}}, \bibnamefont{et~al.}, \href{https://www.nature.com/articles/s41586-021-03853-0}{\bibinfo{title}{Continuous mott transition in semiconductor moir{\'e} superlattices}}, \bibinfo{journal}{Nature} \textbf{\bibinfo{volume}{597}}, \bibinfo{pages}{350} (\bibinfo{year}{2021}{\natexlab{a}}).

\bibitem[{\citenamefont{Xu et~al.}(2022)\citenamefont{Xu, Kang, Watanabe, Taniguchi, Mak, and Shan}}]{xu2022tunable}
\bibinfo{author}{\bibfnamefont{Y.}~\bibnamefont{Xu}}, \bibinfo{author}{\bibfnamefont{K.}~\bibnamefont{Kang}}, \bibinfo{author}{\bibfnamefont{K.}~\bibnamefont{Watanabe}}, \bibinfo{author}{\bibfnamefont{T.}~\bibnamefont{Taniguchi}}, \bibinfo{author}{\bibfnamefont{K.~F.} \bibnamefont{Mak}}, \bibnamefont{and} \bibinfo{author}{\bibfnamefont{J.}~\bibnamefont{Shan}}, \href{https://www.nature.com/articles/s41565-022-01180-7}{\bibinfo{title}{A tunable bilayer hubbard model in twisted wse$_2$}}, \bibinfo{journal}{Nature nanotechnology} \textbf{\bibinfo{volume}{17}}, \bibinfo{pages}{934} (\bibinfo{year}{2022}).

\bibitem[{\citenamefont{Regan et~al.}(2020)\citenamefont{Regan, Wang, Jin, Bakti~Utama, Gao, Wei, Zhao, Zhao, Zhang, Yumigeta et~al.}}]{regan2020mott}
\bibinfo{author}{\bibfnamefont{E.~C.} \bibnamefont{Regan}}, \bibinfo{author}{\bibfnamefont{D.}~\bibnamefont{Wang}}, \bibinfo{author}{\bibfnamefont{C.}~\bibnamefont{Jin}}, \bibinfo{author}{\bibfnamefont{M.~I.} \bibnamefont{Bakti~Utama}}, \bibinfo{author}{\bibfnamefont{B.}~\bibnamefont{Gao}}, \bibinfo{author}{\bibfnamefont{X.}~\bibnamefont{Wei}}, \bibinfo{author}{\bibfnamefont{S.}~\bibnamefont{Zhao}}, \bibinfo{author}{\bibfnamefont{W.}~\bibnamefont{Zhao}}, \bibinfo{author}{\bibfnamefont{Z.}~\bibnamefont{Zhang}}, \bibinfo{author}{\bibfnamefont{K.}~\bibnamefont{Yumigeta}}, \bibnamefont{et~al.}, \href{https://www.nature.com/articles/s41586-020-2092-4}{\bibinfo{title}{Mott and generalized wigner crystal states in wse$_2$/ws$_2$ moir{\'e} superlattices}}, \bibinfo{journal}{Nature} \textbf{\bibinfo{volume}{579}}, \bibinfo{pages}{359} (\bibinfo{year}{2020}).

\bibitem[{\citenamefont{Li et~al.}(2021{\natexlab{b}})\citenamefont{Li, Li, Regan, Wang, Zhao, Kahn, Yumigeta, Blei, Taniguchi, Watanabe et~al.}}]{li2021imaging}
\bibinfo{author}{\bibfnamefont{H.}~\bibnamefont{Li}}, \bibinfo{author}{\bibfnamefont{S.}~\bibnamefont{Li}}, \bibinfo{author}{\bibfnamefont{E.~C.} \bibnamefont{Regan}}, \bibinfo{author}{\bibfnamefont{D.}~\bibnamefont{Wang}}, \bibinfo{author}{\bibfnamefont{W.}~\bibnamefont{Zhao}}, \bibinfo{author}{\bibfnamefont{S.}~\bibnamefont{Kahn}}, \bibinfo{author}{\bibfnamefont{K.}~\bibnamefont{Yumigeta}}, \bibinfo{author}{\bibfnamefont{M.}~\bibnamefont{Blei}}, \bibinfo{author}{\bibfnamefont{T.}~\bibnamefont{Taniguchi}}, \bibinfo{author}{\bibfnamefont{K.}~\bibnamefont{Watanabe}}, \bibnamefont{et~al.}, \href{https://www.nature.com/articles/s41586-021-03874-9}{\bibinfo{title}{Imaging two-dimensional generalized wigner crystals}}, \bibinfo{journal}{Nature} \textbf{\bibinfo{volume}{597}}, \bibinfo{pages}{650} (\bibinfo{year}{2021}{\natexlab{b}}).

\bibitem[{\citenamefont{Jin et~al.}(2021)\citenamefont{Jin, Tao, Li, Xu, Tang, Zhu, Liu, Watanabe, Taniguchi, Hone et~al.}}]{jin2021stripe}
\bibinfo{author}{\bibfnamefont{C.}~\bibnamefont{Jin}}, \bibinfo{author}{\bibfnamefont{Z.}~\bibnamefont{Tao}}, \bibinfo{author}{\bibfnamefont{T.}~\bibnamefont{Li}}, \bibinfo{author}{\bibfnamefont{Y.}~\bibnamefont{Xu}}, \bibinfo{author}{\bibfnamefont{Y.}~\bibnamefont{Tang}}, \bibinfo{author}{\bibfnamefont{J.}~\bibnamefont{Zhu}}, \bibinfo{author}{\bibfnamefont{S.}~\bibnamefont{Liu}}, \bibinfo{author}{\bibfnamefont{K.}~\bibnamefont{Watanabe}}, \bibinfo{author}{\bibfnamefont{T.}~\bibnamefont{Taniguchi}}, \bibinfo{author}{\bibfnamefont{J.~C.} \bibnamefont{Hone}}, \bibnamefont{et~al.}, \href{https://www.nature.com/articles/s41563-021-00959-8}{\bibinfo{title}{Stripe phases in wse2/ws2 moir{\'e} superlattices}}, \bibinfo{journal}{Nature Materials} \textbf{\bibinfo{volume}{20}}, \bibinfo{pages}{940} (\bibinfo{year}{2021}).

\bibitem[{\citenamefont{Li et~al.}(2021{\natexlab{c}})\citenamefont{Li, Jiang, Shen, Zhang, Li, Tao, Devakul, Watanabe, Taniguchi, Fu et~al.}}]{li2021quantum}
\bibinfo{author}{\bibfnamefont{T.}~\bibnamefont{Li}}, \bibinfo{author}{\bibfnamefont{S.}~\bibnamefont{Jiang}}, \bibinfo{author}{\bibfnamefont{B.}~\bibnamefont{Shen}}, \bibinfo{author}{\bibfnamefont{Y.}~\bibnamefont{Zhang}}, \bibinfo{author}{\bibfnamefont{L.}~\bibnamefont{Li}}, \bibinfo{author}{\bibfnamefont{Z.}~\bibnamefont{Tao}}, \bibinfo{author}{\bibfnamefont{T.}~\bibnamefont{Devakul}}, \bibinfo{author}{\bibfnamefont{K.}~\bibnamefont{Watanabe}}, \bibinfo{author}{\bibfnamefont{T.}~\bibnamefont{Taniguchi}}, \bibinfo{author}{\bibfnamefont{L.}~\bibnamefont{Fu}}, \bibnamefont{et~al.}, \href{https://www.nature.com/articles/s41586-021-04171-1}{\bibinfo{title}{Quantum anomalous hall effect from intertwined moir{\'e} bands}}, \bibinfo{journal}{Nature} \textbf{\bibinfo{volume}{600}}, \bibinfo{pages}{641} (\bibinfo{year}{2021}{\natexlab{c}}).

\bibitem[{\citenamefont{Park et~al.}(2023)\citenamefont{Park, Cai, Anderson, Zhang, Zhu, Liu, Wang, Holtzmann, Hu, Liu et~al.}}]{park2023observation}
\bibinfo{author}{\bibfnamefont{H.}~\bibnamefont{Park}}, \bibinfo{author}{\bibfnamefont{J.}~\bibnamefont{Cai}}, \bibinfo{author}{\bibfnamefont{E.}~\bibnamefont{Anderson}}, \bibinfo{author}{\bibfnamefont{Y.}~\bibnamefont{Zhang}}, \bibinfo{author}{\bibfnamefont{J.}~\bibnamefont{Zhu}}, \bibinfo{author}{\bibfnamefont{X.}~\bibnamefont{Liu}}, \bibinfo{author}{\bibfnamefont{C.}~\bibnamefont{Wang}}, \bibinfo{author}{\bibfnamefont{W.}~\bibnamefont{Holtzmann}}, \bibinfo{author}{\bibfnamefont{C.}~\bibnamefont{Hu}}, \bibinfo{author}{\bibfnamefont{Z.}~\bibnamefont{Liu}}, \bibnamefont{et~al.}, \href{https://www.nature.com/articles/s41586-023-06536-0}{\bibinfo{title}{Observation of fractionally quantized anomalous hall effect}}, \bibinfo{journal}{Nature} \textbf{\bibinfo{volume}{622}}, \bibinfo{pages}{74} (\bibinfo{year}{2023}).

\bibitem[{\citenamefont{Xu et~al.}(2023)\citenamefont{Xu, Sun, Jia, Liu, Xu, Li, Gu, Watanabe, Taniguchi, Tong et~al.}}]{PhysRevX.13.031037}
\bibinfo{author}{\bibfnamefont{F.}~\bibnamefont{Xu}}, \bibinfo{author}{\bibfnamefont{Z.}~\bibnamefont{Sun}}, \bibinfo{author}{\bibfnamefont{T.}~\bibnamefont{Jia}}, \bibinfo{author}{\bibfnamefont{C.}~\bibnamefont{Liu}}, \bibinfo{author}{\bibfnamefont{C.}~\bibnamefont{Xu}}, \bibinfo{author}{\bibfnamefont{C.}~\bibnamefont{Li}}, \bibinfo{author}{\bibfnamefont{Y.}~\bibnamefont{Gu}}, \bibinfo{author}{\bibfnamefont{K.}~\bibnamefont{Watanabe}}, \bibinfo{author}{\bibfnamefont{T.}~\bibnamefont{Taniguchi}}, \bibinfo{author}{\bibfnamefont{B.}~\bibnamefont{Tong}}, \bibnamefont{et~al.}, \href{https://link.aps.org/doi/10.1103/PhysRevX.13.031037}{\bibinfo{title}{Observation of integer and fractional quantum anomalous hall effects in twisted bilayer ${\mathrm{mote}}_{2}$}}, \bibinfo{journal}{Phys. Rev. X} \textbf{\bibinfo{volume}{13}}, \bibinfo{pages}{031037} (\bibinfo{year}{2023}).

\bibitem[{\citenamefont{Kang et~al.}(2024)\citenamefont{Kang, Shen, Qiu, Zeng, Xia, Watanabe, Taniguchi, Shan, and Mak}}]{kang2024evidence}
\bibinfo{author}{\bibfnamefont{K.}~\bibnamefont{Kang}}, \bibinfo{author}{\bibfnamefont{B.}~\bibnamefont{Shen}}, \bibinfo{author}{\bibfnamefont{Y.}~\bibnamefont{Qiu}}, \bibinfo{author}{\bibfnamefont{Y.}~\bibnamefont{Zeng}}, \bibinfo{author}{\bibfnamefont{Z.}~\bibnamefont{Xia}}, \bibinfo{author}{\bibfnamefont{K.}~\bibnamefont{Watanabe}}, \bibinfo{author}{\bibfnamefont{T.}~\bibnamefont{Taniguchi}}, \bibinfo{author}{\bibfnamefont{J.}~\bibnamefont{Shan}}, \bibnamefont{and} \bibinfo{author}{\bibfnamefont{K.~F.} \bibnamefont{Mak}}, \href{https://www.nature.com/articles/s41586-024-07214-5}{\bibinfo{title}{Evidence of the fractional quantum spin hall effect in moir{\'e} mote2}}, \bibinfo{journal}{Nature} \textbf{\bibinfo{volume}{628}}, \bibinfo{pages}{522} (\bibinfo{year}{2024}).

\bibitem[{\citenamefont{Xia et~al.}(2025)\citenamefont{Xia, Han, Watanabe, Taniguchi, Shan, and Mak}}]{xia2025superconductivity}
\bibinfo{author}{\bibfnamefont{Y.}~\bibnamefont{Xia}}, \bibinfo{author}{\bibfnamefont{Z.}~\bibnamefont{Han}}, \bibinfo{author}{\bibfnamefont{K.}~\bibnamefont{Watanabe}}, \bibinfo{author}{\bibfnamefont{T.}~\bibnamefont{Taniguchi}}, \bibinfo{author}{\bibfnamefont{J.}~\bibnamefont{Shan}}, \bibnamefont{and} \bibinfo{author}{\bibfnamefont{K.~F.} \bibnamefont{Mak}}, \href{https://www.nature.com/articles/s41586-024-08116-2}{\bibinfo{title}{Superconductivity in twisted bilayer wse2}}, \bibinfo{journal}{Nature} \textbf{\bibinfo{volume}{637}}, \bibinfo{pages}{833} (\bibinfo{year}{2025}).

\bibitem[{\citenamefont{Guo et~al.}(2025)\citenamefont{Guo, Pack, Swann, Holtzman, Cothrine, Watanabe, Taniguchi, Mandrus, Barmak, Hone et~al.}}]{guo2025superconductivity}
\bibinfo{author}{\bibfnamefont{Y.}~\bibnamefont{Guo}}, \bibinfo{author}{\bibfnamefont{J.}~\bibnamefont{Pack}}, \bibinfo{author}{\bibfnamefont{J.}~\bibnamefont{Swann}}, \bibinfo{author}{\bibfnamefont{L.}~\bibnamefont{Holtzman}}, \bibinfo{author}{\bibfnamefont{M.}~\bibnamefont{Cothrine}}, \bibinfo{author}{\bibfnamefont{K.}~\bibnamefont{Watanabe}}, \bibinfo{author}{\bibfnamefont{T.}~\bibnamefont{Taniguchi}}, \bibinfo{author}{\bibfnamefont{D.~G.} \bibnamefont{Mandrus}}, \bibinfo{author}{\bibfnamefont{K.}~\bibnamefont{Barmak}}, \bibinfo{author}{\bibfnamefont{J.}~\bibnamefont{Hone}}, \bibnamefont{et~al.}, \href{https://www.nature.com/articles/s41586-024-08381-1}{\bibinfo{title}{Superconductivity in 5.0° twisted bilayer wse2}}, \bibinfo{journal}{Nature} \textbf{\bibinfo{volume}{637}}, \bibinfo{pages}{839} (\bibinfo{year}{2025}).

\bibitem[{\citenamefont{Schrade and Fu}(2024)}]{PhysRevB.110.035143}
\bibinfo{author}{\bibfnamefont{C.}~\bibnamefont{Schrade}} \bibnamefont{and} \bibinfo{author}{\bibfnamefont{L.}~\bibnamefont{Fu}}, \href{https://link.aps.org/doi/10.1103/PhysRevB.110.035143}{\bibinfo{title}{Nematic, chiral, and topological superconductivity in twisted transition metal dichalcogenides}}, \bibinfo{journal}{Phys. Rev. B} \textbf{\bibinfo{volume}{110}}, \bibinfo{pages}{035143} (\bibinfo{year}{2024}).

\bibitem[{\citenamefont{Chubukov and Varma}(2025)}]{PhysRevB.111.014507}
\bibinfo{author}{\bibfnamefont{A.~V.} \bibnamefont{Chubukov}} \bibnamefont{and} \bibinfo{author}{\bibfnamefont{C.~M.} \bibnamefont{Varma}}, \href{https://link.aps.org/doi/10.1103/PhysRevB.111.014507}{\bibinfo{title}{Quantum criticality and superconductivity in twisted transition metal dichalcogenides}}, \bibinfo{journal}{Phys. Rev. B} \textbf{\bibinfo{volume}{111}}, \bibinfo{pages}{014507} (\bibinfo{year}{2025}).

\bibitem[{\citenamefont{Zhu et~al.}(2025{\natexlab{a}})\citenamefont{Zhu, Chou, Xie, and Das~Sarma}}]{PhysRevB.111.L060501}
\bibinfo{author}{\bibfnamefont{J.}~\bibnamefont{Zhu}}, \bibinfo{author}{\bibfnamefont{Y.-Z.} \bibnamefont{Chou}}, \bibinfo{author}{\bibfnamefont{M.}~\bibnamefont{Xie}}, \bibnamefont{and} \bibinfo{author}{\bibfnamefont{S.}~\bibnamefont{Das~Sarma}}, \href{https://link.aps.org/doi/10.1103/PhysRevB.111.L060501}{\bibinfo{title}{Superconductivity in twisted transition metal dichalcogenide homobilayers}}, \bibinfo{journal}{Phys. Rev. B} \textbf{\bibinfo{volume}{111}}, \bibinfo{pages}{L060501} (\bibinfo{year}{2025}{\natexlab{a}}).

\bibitem[{\citenamefont{Christos et~al.}(2024)\citenamefont{Christos, Bonetti, and Scheurer}}]{christos2024approximate}
\bibinfo{author}{\bibfnamefont{M.}~\bibnamefont{Christos}}, \bibinfo{author}{\bibfnamefont{P.~M.} \bibnamefont{Bonetti}}, \bibnamefont{and} \bibinfo{author}{\bibfnamefont{M.~S.} \bibnamefont{Scheurer}}, \href{https://arxiv.org/abs/2407.02393}{\bibinfo{title}{Approximate symmetries, insulators, and superconductivity in continuum-model description of twisted wse$_2$}}, \bibinfo{journal}{arXiv preprint arXiv:2407.02393}  (\bibinfo{year}{2024}).

\bibitem[{\citenamefont{Guerci et~al.}(2024)\citenamefont{Guerci, Kaplan, Ingham, Pixley, and Millis}}]{guerci2024topological}
\bibinfo{author}{\bibfnamefont{D.}~\bibnamefont{Guerci}}, \bibinfo{author}{\bibfnamefont{D.}~\bibnamefont{Kaplan}}, \bibinfo{author}{\bibfnamefont{J.}~\bibnamefont{Ingham}}, \bibinfo{author}{\bibfnamefont{J.}~\bibnamefont{Pixley}}, \bibnamefont{and} \bibinfo{author}{\bibfnamefont{A.~J.} \bibnamefont{Millis}}, \href{https://arxiv.org/abs/2408.16075}{\bibinfo{title}{Topological superconductivity from repulsive interactions in twisted wse$_2$}}, \bibinfo{journal}{arXiv preprint arXiv:2408.16075}  (\bibinfo{year}{2024}).

\bibitem[{\citenamefont{Tuo et~al.}(2024)\citenamefont{Tuo, Li, Wu, Sun, and Yao}}]{tuo2024theory}
\bibinfo{author}{\bibfnamefont{C.}~\bibnamefont{Tuo}}, \bibinfo{author}{\bibfnamefont{M.-R.} \bibnamefont{Li}}, \bibinfo{author}{\bibfnamefont{Z.}~\bibnamefont{Wu}}, \bibinfo{author}{\bibfnamefont{W.}~\bibnamefont{Sun}}, \bibnamefont{and} \bibinfo{author}{\bibfnamefont{H.}~\bibnamefont{Yao}}, \href{https://arxiv.org/abs/2409.06779}{\bibinfo{title}{Theory of topological superconductivity and antiferromagnetic correlated insulators in twisted bilayer wse$_2$}}, \bibinfo{journal}{arXiv preprint arXiv:2409.06779}  (\bibinfo{year}{2024}).

\bibitem[{\citenamefont{Qin et~al.}(2024)\citenamefont{Qin, Qiu, and Wu}}]{qin2024kohn}
\bibinfo{author}{\bibfnamefont{W.}~\bibnamefont{Qin}}, \bibinfo{author}{\bibfnamefont{W.-X.} \bibnamefont{Qiu}}, \bibnamefont{and} \bibinfo{author}{\bibfnamefont{F.}~\bibnamefont{Wu}}, \href{https://arxiv.org/abs/2409.16114}{\bibinfo{title}{Kohn-luttinger mechanism of superconductivity in twisted bilayer wse$_2$: Gate-tunable unconventional pairing symmetry}}, \bibinfo{journal}{arXiv preprint arXiv:2409.16114}  (\bibinfo{year}{2024}).

\bibitem[{\citenamefont{Fischer et~al.}(2024)\citenamefont{Fischer, Klebl, Cr{\'e}pel, Ryee, Rubio, Xian, Wehling, Georges, Kennes, and Millis}}]{fischer2024theory}
\bibinfo{author}{\bibfnamefont{A.}~\bibnamefont{Fischer}}, \bibinfo{author}{\bibfnamefont{L.}~\bibnamefont{Klebl}}, \bibinfo{author}{\bibfnamefont{V.}~\bibnamefont{Cr{\'e}pel}}, \bibinfo{author}{\bibfnamefont{S.}~\bibnamefont{Ryee}}, \bibinfo{author}{\bibfnamefont{A.}~\bibnamefont{Rubio}}, \bibinfo{author}{\bibfnamefont{L.}~\bibnamefont{Xian}}, \bibinfo{author}{\bibfnamefont{T.~O.} \bibnamefont{Wehling}}, \bibinfo{author}{\bibfnamefont{A.}~\bibnamefont{Georges}}, \bibinfo{author}{\bibfnamefont{D.~M.} \bibnamefont{Kennes}}, \bibnamefont{and} \bibinfo{author}{\bibfnamefont{A.~J.} \bibnamefont{Millis}}, \href{https://www.arxiv.org/abs/2412.14296}{\bibinfo{title}{Theory of intervalley-coherent afm order and topological superconductivity in twse$_2$}}, \bibinfo{journal}{arXiv preprint arXiv:2412.14296}  (\bibinfo{year}{2024}).

\bibitem[{\citenamefont{Wietek et~al.}(2022)\citenamefont{Wietek, Wang, Zang, Cano, Georges, and Millis}}]{PhysRevResearch.4.043048}
\bibinfo{author}{\bibfnamefont{A.}~\bibnamefont{Wietek}}, \bibinfo{author}{\bibfnamefont{J.}~\bibnamefont{Wang}}, \bibinfo{author}{\bibfnamefont{J.}~\bibnamefont{Zang}}, \bibinfo{author}{\bibfnamefont{J.}~\bibnamefont{Cano}}, \bibinfo{author}{\bibfnamefont{A.}~\bibnamefont{Georges}}, \bibnamefont{and} \bibinfo{author}{\bibfnamefont{A.}~\bibnamefont{Millis}}, \href{https://link.aps.org/doi/10.1103/PhysRevResearch.4.043048}{\bibinfo{title}{Tunable stripe order and weak superconductivity in the moir\'e hubbard model}}, \bibinfo{journal}{Phys. Rev. Res.} \textbf{\bibinfo{volume}{4}}, \bibinfo{pages}{043048} (\bibinfo{year}{2022}).

\bibitem[{\citenamefont{Klebl et~al.}(2023)\citenamefont{Klebl, Fischer, Classen, Scherer, and Kennes}}]{PhysRevResearch.5.L012034}
\bibinfo{author}{\bibfnamefont{L.}~\bibnamefont{Klebl}}, \bibinfo{author}{\bibfnamefont{A.}~\bibnamefont{Fischer}}, \bibinfo{author}{\bibfnamefont{L.}~\bibnamefont{Classen}}, \bibinfo{author}{\bibfnamefont{M.~M.} \bibnamefont{Scherer}}, \bibnamefont{and} \bibinfo{author}{\bibfnamefont{D.~M.} \bibnamefont{Kennes}}, \href{https://link.aps.org/doi/10.1103/PhysRevResearch.5.L012034}{\bibinfo{title}{Competition of density waves and superconductivity in twisted tungsten diselenide}}, \bibinfo{journal}{Phys. Rev. Res.} \textbf{\bibinfo{volume}{5}}, \bibinfo{pages}{L012034} (\bibinfo{year}{2023}).

\bibitem[{\citenamefont{Xie et~al.}(2025)\citenamefont{Xie, Chen, Sur, Fang, Cano, and Si}}]{PhysRevLett.134.136503}
\bibinfo{author}{\bibfnamefont{F.}~\bibnamefont{Xie}}, \bibinfo{author}{\bibfnamefont{L.}~\bibnamefont{Chen}}, \bibinfo{author}{\bibfnamefont{S.}~\bibnamefont{Sur}}, \bibinfo{author}{\bibfnamefont{Y.}~\bibnamefont{Fang}}, \bibinfo{author}{\bibfnamefont{J.}~\bibnamefont{Cano}}, \bibnamefont{and} \bibinfo{author}{\bibfnamefont{Q.}~\bibnamefont{Si}}, \href{https://link.aps.org/doi/10.1103/PhysRevLett.134.136503}{\bibinfo{title}{Superconductivity in twisted wse$_2$ from topology-induced quantum fluctuations}}, \bibinfo{journal}{Phys. Rev. Lett.} \textbf{\bibinfo{volume}{134}}, \bibinfo{pages}{136503} (\bibinfo{year}{2025}).

\bibitem[{\citenamefont{Kim et~al.}(2025)\citenamefont{Kim, Mendez-Valderrama, Wang, and Chowdhury}}]{kim2025theory}
\bibinfo{author}{\bibfnamefont{S.}~\bibnamefont{Kim}}, \bibinfo{author}{\bibfnamefont{J.~F.} \bibnamefont{Mendez-Valderrama}}, \bibinfo{author}{\bibfnamefont{X.}~\bibnamefont{Wang}}, \bibnamefont{and} \bibinfo{author}{\bibfnamefont{D.}~\bibnamefont{Chowdhury}}, \href{https://www.nature.com/articles/s41467-025-56816-8}{\bibinfo{title}{Theory of correlated insulators and superconductor at $\nu$= 1 in twisted wse$_2$}}, \bibinfo{journal}{Nature Communications} \textbf{\bibinfo{volume}{16}}, \bibinfo{pages}{1701} (\bibinfo{year}{2025}).

\bibitem[{\citenamefont{Zhu et~al.}(2025{\natexlab{b}})\citenamefont{Zhu, Chou, Huang, and Sarma}}]{zhu2025plane}
\bibinfo{author}{\bibfnamefont{J.}~\bibnamefont{Zhu}}, \bibinfo{author}{\bibfnamefont{Y.-Z.} \bibnamefont{Chou}}, \bibinfo{author}{\bibfnamefont{Y.}~\bibnamefont{Huang}}, \bibnamefont{and} \bibinfo{author}{\bibfnamefont{S.~D.} \bibnamefont{Sarma}}, \href{https://arxiv.org/abs/2503.18946}{\bibinfo{title}{In-plane magnetic field-induced orbital fflo superconductivity in twisted wse$_2 $ homobilayers}}, \bibinfo{journal}{arXiv preprint arXiv:2503.18946}  (\bibinfo{year}{2025}{\natexlab{b}}).

\bibitem[{\citenamefont{Xu et~al.}(2025{\natexlab{a}})\citenamefont{Xu, Sun, Li, Zheng, Xu, Gao, Jia, Watanabe, Taniguchi, Tong et~al.}}]{xu2025signatures}
\bibinfo{author}{\bibfnamefont{F.}~\bibnamefont{Xu}}, \bibinfo{author}{\bibfnamefont{Z.}~\bibnamefont{Sun}}, \bibinfo{author}{\bibfnamefont{J.}~\bibnamefont{Li}}, \bibinfo{author}{\bibfnamefont{C.}~\bibnamefont{Zheng}}, \bibinfo{author}{\bibfnamefont{C.}~\bibnamefont{Xu}}, \bibinfo{author}{\bibfnamefont{J.}~\bibnamefont{Gao}}, \bibinfo{author}{\bibfnamefont{T.}~\bibnamefont{Jia}}, \bibinfo{author}{\bibfnamefont{K.}~\bibnamefont{Watanabe}}, \bibinfo{author}{\bibfnamefont{T.}~\bibnamefont{Taniguchi}}, \bibinfo{author}{\bibfnamefont{B.}~\bibnamefont{Tong}}, \bibnamefont{et~al.}, \href{https://arxiv.org/abs/2504.06972}{\bibinfo{title}{Signatures of unconventional superconductivity near reentrant and fractional quantum anomalous hall insulators}}, \bibinfo{journal}{arXiv preprint arXiv:2504.06972}  (\bibinfo{year}{2025}{\natexlab{a}}).

\bibitem[{\citenamefont{Xu et~al.}(2025{\natexlab{b}})\citenamefont{Xu, Zou, Peshcherenko, Jahin, Li, Lin, and Zhang}}]{xu2025chiral}
\bibinfo{author}{\bibfnamefont{C.}~\bibnamefont{Xu}}, \bibinfo{author}{\bibfnamefont{N.}~\bibnamefont{Zou}}, \bibinfo{author}{\bibfnamefont{N.}~\bibnamefont{Peshcherenko}}, \bibinfo{author}{\bibfnamefont{A.}~\bibnamefont{Jahin}}, \bibinfo{author}{\bibfnamefont{T.}~\bibnamefont{Li}}, \bibinfo{author}{\bibfnamefont{S.-Z.} \bibnamefont{Lin}}, \bibnamefont{and} \bibinfo{author}{\bibfnamefont{Y.}~\bibnamefont{Zhang}}, \href{https://arxiv.org/abs/2504.07082}{\bibinfo{title}{Chiral superconductivity from spin polarized chern band in twisted mote $ \_2$}}, \bibinfo{journal}{arXiv preprint arXiv:2504.07082}  (\bibinfo{year}{2025}{\natexlab{b}}).

\bibitem[{\citenamefont{Shi and Senthil}(2025)}]{shi2025}
\bibinfo{author}{\bibfnamefont{Z.~D.} \bibnamefont{Shi}} \bibnamefont{and} \bibinfo{author}{\bibfnamefont{T.}~\bibnamefont{Senthil}}, \href{https://arxiv.org/abs/2506.02128}{\bibinfo{title}{Anyon delocalization transitions out of a disordered fqah insulator}} (\bibinfo{year}{2025}), \eprint{2506.02128}.

\bibitem[{\citenamefont{Nosov et~al.}(2025)\citenamefont{Nosov, Han, and Khalaf}}]{nosov2025}
\bibinfo{author}{\bibfnamefont{P.~A.} \bibnamefont{Nosov}}, \bibinfo{author}{\bibfnamefont{Z.}~\bibnamefont{Han}}, \bibnamefont{and} \bibinfo{author}{\bibfnamefont{E.}~\bibnamefont{Khalaf}}, \href{https://arxiv.org/abs/2506.02108}{\bibinfo{title}{Anyon superconductivity and plateau transitions in doped fractional quantum anomalous hall insulators}} (\bibinfo{year}{2025}), \eprint{2506.02108}.

\bibitem[{\citenamefont{Huang et~al.}(2025)\citenamefont{Huang, Musser, Zhu, Chou, and Sarma}}]{huang2025}
\bibinfo{author}{\bibfnamefont{Y.}~\bibnamefont{Huang}}, \bibinfo{author}{\bibfnamefont{S.}~\bibnamefont{Musser}}, \bibinfo{author}{\bibfnamefont{J.}~\bibnamefont{Zhu}}, \bibinfo{author}{\bibfnamefont{Y.-Z.} \bibnamefont{Chou}}, \bibnamefont{and} \bibinfo{author}{\bibfnamefont{S.~D.} \bibnamefont{Sarma}}, \href{https://arxiv.org/abs/2506.10965}{\bibinfo{title}{Apparent inconsistency between streda formula and hall conductivity in reentrant integer quantum anomalous hall effect in twisted mote$_2$}} (\bibinfo{year}{2025}), \eprint{2506.10965}.

\bibitem[{\citenamefont{Guerci et~al.}(2025)\citenamefont{Guerci, Abouelkomsan, and Fu}}]{guerci2025}
\bibinfo{author}{\bibfnamefont{D.}~\bibnamefont{Guerci}}, \bibinfo{author}{\bibfnamefont{A.}~\bibnamefont{Abouelkomsan}}, \bibnamefont{and} \bibinfo{author}{\bibfnamefont{L.}~\bibnamefont{Fu}}, \href{https://arxiv.org/abs/2506.10938}{\bibinfo{title}{From fractionalization to chiral topological superconductivity in flat chern band}} (\bibinfo{year}{2025}), \eprint{2506.10938}.

\bibitem[{\citenamefont{Zhang}(2025)}]{zhang2025}
\bibinfo{author}{\bibfnamefont{Y.-H.} \bibnamefont{Zhang}}, \href{https://arxiv.org/abs/2506.00110}{\bibinfo{title}{Holon metal, charge-density-wave and chiral superconductor from doping fractional chern insulator and su(3)$_1$ chiral spin liquid}} (\bibinfo{year}{2025}), \eprint{2506.00110}.

\bibitem[{\citenamefont{Hu et~al.}(2025)\citenamefont{Hu, Daido, Sun, Xie, and Law}}]{hu2025layer}
\bibinfo{author}{\bibfnamefont{J.-X.} \bibnamefont{Hu}}, \bibinfo{author}{\bibfnamefont{A.}~\bibnamefont{Daido}}, \bibinfo{author}{\bibfnamefont{Z.-T.} \bibnamefont{Sun}}, \bibinfo{author}{\bibfnamefont{Y.-M.} \bibnamefont{Xie}}, \bibnamefont{and} \bibinfo{author}{\bibfnamefont{K.}~\bibnamefont{Law}}, \href{https://arxiv.org/abs/2506.12767}{\bibinfo{title}{Layer pseudospin superconductivity in twisted mote$_2$}}, \bibinfo{journal}{arXiv preprint arXiv:2506.12767}  (\bibinfo{year}{2025}).

\bibitem[{\citenamefont{Zhang et~al.}(2024{\natexlab{a}})\citenamefont{Zhang, Wang, Liu, Fan, Cao, and Xiao}}]{zhang2024polarization}
\bibinfo{author}{\bibfnamefont{X.-W.} \bibnamefont{Zhang}}, \bibinfo{author}{\bibfnamefont{C.}~\bibnamefont{Wang}}, \bibinfo{author}{\bibfnamefont{X.}~\bibnamefont{Liu}}, \bibinfo{author}{\bibfnamefont{Y.}~\bibnamefont{Fan}}, \bibinfo{author}{\bibfnamefont{T.}~\bibnamefont{Cao}}, \bibnamefont{and} \bibinfo{author}{\bibfnamefont{D.}~\bibnamefont{Xiao}}, \href{https://www.nature.com/articles/s41467-024-48511-x}{\bibinfo{title}{Polarization-driven band topology evolution in twisted mote2 and wse2}}, \bibinfo{journal}{Nature Communications} \textbf{\bibinfo{volume}{15}}, \bibinfo{pages}{4223} (\bibinfo{year}{2024}{\natexlab{a}}).

\bibitem[{\citenamefont{Wu et~al.}(2019)\citenamefont{Wu, Lovorn, Tutuc, Martin, and MacDonald}}]{WuFengcheng2019}
\bibinfo{author}{\bibfnamefont{F.}~\bibnamefont{Wu}}, \bibinfo{author}{\bibfnamefont{T.}~\bibnamefont{Lovorn}}, \bibinfo{author}{\bibfnamefont{E.}~\bibnamefont{Tutuc}}, \bibinfo{author}{\bibfnamefont{I.}~\bibnamefont{Martin}}, \bibnamefont{and} \bibinfo{author}{\bibfnamefont{A.~H.} \bibnamefont{MacDonald}}, \href{https://link.aps.org/doi/10.1103/PhysRevLett.122.086402}{\bibinfo{title}{Topological insulators in twisted transition metal dichalcogenide homobilayers}}, \bibinfo{journal}{Phys. Rev. Lett.} \textbf{\bibinfo{volume}{122}}, \bibinfo{pages}{086402} (\bibinfo{year}{2019}).

\bibitem[{\citenamefont{Xu et~al.}(2024)\citenamefont{Xu, Li, Xu, Bi, and Zhang}}]{xu2024maximally}
\bibinfo{author}{\bibfnamefont{C.}~\bibnamefont{Xu}}, \bibinfo{author}{\bibfnamefont{J.}~\bibnamefont{Li}}, \bibinfo{author}{\bibfnamefont{Y.}~\bibnamefont{Xu}}, \bibinfo{author}{\bibfnamefont{Z.}~\bibnamefont{Bi}}, \bibnamefont{and} \bibinfo{author}{\bibfnamefont{Y.}~\bibnamefont{Zhang}}, \href{https://www.pnas.org/doi/abs/10.1073/pnas.2316749121}{\bibinfo{title}{Maximally localized wannier functions, interaction models, and fractional quantum anomalous hall effect in twisted bilayer mote2}}, \bibinfo{journal}{Proceedings of the National Academy of Sciences} \textbf{\bibinfo{volume}{121}}, \bibinfo{pages}{e2316749121} (\bibinfo{year}{2024}).

\bibitem[{\citenamefont{Mao et~al.}(2024)\citenamefont{Mao, Xu, Li, Bao, Liu, Xu, Felser, Fu, and Zhang}}]{mao2024transfer}
\bibinfo{author}{\bibfnamefont{N.}~\bibnamefont{Mao}}, \bibinfo{author}{\bibfnamefont{C.}~\bibnamefont{Xu}}, \bibinfo{author}{\bibfnamefont{J.}~\bibnamefont{Li}}, \bibinfo{author}{\bibfnamefont{T.}~\bibnamefont{Bao}}, \bibinfo{author}{\bibfnamefont{P.}~\bibnamefont{Liu}}, \bibinfo{author}{\bibfnamefont{Y.}~\bibnamefont{Xu}}, \bibinfo{author}{\bibfnamefont{C.}~\bibnamefont{Felser}}, \bibinfo{author}{\bibfnamefont{L.}~\bibnamefont{Fu}}, \bibnamefont{and} \bibinfo{author}{\bibfnamefont{Y.}~\bibnamefont{Zhang}}, \href{https://www.nature.com/articles/s42005-024-01754-y}{\bibinfo{title}{Transfer learning relaxation, electronic structure and continuum model for twisted bilayer mote2}}, \bibinfo{journal}{Communications Physics} \textbf{\bibinfo{volume}{7}}, \bibinfo{pages}{262} (\bibinfo{year}{2024}).

\bibitem[{\citenamefont{Jia et~al.}(2024)\citenamefont{Jia, Yu, Liu, Herzog-Arbeitman, Qi, Pi, Regnault, Weng, Bernevig, and Wu}}]{Yujin2024}
\bibinfo{author}{\bibfnamefont{Y.}~\bibnamefont{Jia}}, \bibinfo{author}{\bibfnamefont{J.}~\bibnamefont{Yu}}, \bibinfo{author}{\bibfnamefont{J.}~\bibnamefont{Liu}}, \bibinfo{author}{\bibfnamefont{J.}~\bibnamefont{Herzog-Arbeitman}}, \bibinfo{author}{\bibfnamefont{Z.}~\bibnamefont{Qi}}, \bibinfo{author}{\bibfnamefont{H.}~\bibnamefont{Pi}}, \bibinfo{author}{\bibfnamefont{N.}~\bibnamefont{Regnault}}, \bibinfo{author}{\bibfnamefont{H.}~\bibnamefont{Weng}}, \bibinfo{author}{\bibfnamefont{B.~A.} \bibnamefont{Bernevig}}, \bibnamefont{and} \bibinfo{author}{\bibfnamefont{Q.}~\bibnamefont{Wu}}, \href{https://link.aps.org/doi/10.1103/PhysRevB.109.205121}{\bibinfo{title}{Moir\'e fractional chern insulators. i. first-principles calculations and continuum models of twisted bilayer ${\mathrm{mote}}_{2}$}}, \bibinfo{journal}{Phys. Rev. B} \textbf{\bibinfo{volume}{109}}, \bibinfo{pages}{205121} (\bibinfo{year}{2024}).

\bibitem[{\citenamefont{Kohn and Luttinger}(1965)}]{kohn1965new}
\bibinfo{author}{\bibfnamefont{W.}~\bibnamefont{Kohn}} \bibnamefont{and} \bibinfo{author}{\bibfnamefont{J.}~\bibnamefont{Luttinger}}, \href{https://journals.aps.org/prl/abstract/10.1103/PhysRevLett.15.524}{\bibinfo{title}{New mechanism for superconductivity}}, \bibinfo{journal}{Physical Review Letters} \textbf{\bibinfo{volume}{15}}, \bibinfo{pages}{524} (\bibinfo{year}{1965}).

\bibitem[{\citenamefont{Chubukov}(1993)}]{chubukov1993kohn}
\bibinfo{author}{\bibfnamefont{A.~V.} \bibnamefont{Chubukov}}, \href{https://journals.aps.org/prb/abstract/10.1103/PhysRevB.48.1097}{\bibinfo{title}{Kohn-luttinger effect and the instability of a two-dimensional repulsive fermi liquid at t= 0}}, \bibinfo{journal}{Physical Review B} \textbf{\bibinfo{volume}{48}}, \bibinfo{pages}{1097} (\bibinfo{year}{1993}).

\bibitem[{\citenamefont{Davydova et~al.}(2024)\citenamefont{Davydova, Geier, and Fu}}]{davydovageier2024}
\bibinfo{author}{\bibfnamefont{M.}~\bibnamefont{Davydova}}, \bibinfo{author}{\bibfnamefont{M.}~\bibnamefont{Geier}}, \bibnamefont{and} \bibinfo{author}{\bibfnamefont{L.}~\bibnamefont{Fu}}, \href{https://www.science.org/doi/abs/10.1126/sciadv.adr4817}{\bibinfo{title}{Nonreciprocal superconductivity}}, \bibinfo{journal}{Science Advances} \textbf{\bibinfo{volume}{10}}, \bibinfo{pages}{eadr4817} (\bibinfo{year}{2024}), \eprint{https://www.science.org/doi/pdf/10.1126/sciadv.adr4817}.

\bibitem[{\citenamefont{Chen et~al.}(2025)\citenamefont{Chen, Scheurer, and Schrade}}]{chen2025}
\bibinfo{author}{\bibfnamefont{Y.}~\bibnamefont{Chen}}, \bibinfo{author}{\bibfnamefont{M.~S.} \bibnamefont{Scheurer}}, \bibnamefont{and} \bibinfo{author}{\bibfnamefont{C.}~\bibnamefont{Schrade}}, \href{https://arxiv.org/abs/2503.16391}{\bibinfo{title}{Intrinsic superconducting diode effect and nonreciprocal superconductivity in rhombohedral graphene multilayers}} (\bibinfo{year}{2025}), \eprint{2503.16391}, \urlprefix\url{https://arxiv.org/abs/2503.16391}.

\bibitem[{\citenamefont{de~Azambuja and Möckli}(2025)}]{deazambuja2025}
\bibinfo{author}{\bibfnamefont{M.~K.} \bibnamefont{de~Azambuja}} \bibnamefont{and} \bibinfo{author}{\bibfnamefont{D.}~\bibnamefont{Möckli}}, \href{https://arxiv.org/abs/2506.05550}{\bibinfo{title}{Axionic nonreciprocal superconductivity}} (\bibinfo{year}{2025}), \eprint{2506.05550}.

\bibitem[{\citenamefont{Nadeem et~al.}(2023)\citenamefont{Nadeem, Fuhrer, and Wang}}]{nadeem2023superconducting}
\bibinfo{author}{\bibfnamefont{M.}~\bibnamefont{Nadeem}}, \bibinfo{author}{\bibfnamefont{M.~S.} \bibnamefont{Fuhrer}}, \bibnamefont{and} \bibinfo{author}{\bibfnamefont{X.}~\bibnamefont{Wang}}, \href{https://www.nature.com/articles/s42254-023-00632-w}{\bibinfo{title}{The superconducting diode effect}}, \bibinfo{journal}{Nature Reviews Physics} \textbf{\bibinfo{volume}{5}}, \bibinfo{pages}{558} (\bibinfo{year}{2023}).

\bibitem[{\citenamefont{Ando et~al.}(2020)\citenamefont{Ando, Miyasaka, Li, Ishizuka, Arakawa, Shiota, Moriyama, Yanase, and Ono}}]{ando_observation_2020}
\bibinfo{author}{\bibfnamefont{F.}~\bibnamefont{Ando}}, \bibinfo{author}{\bibfnamefont{Y.}~\bibnamefont{Miyasaka}}, \bibinfo{author}{\bibfnamefont{T.}~\bibnamefont{Li}}, \bibinfo{author}{\bibfnamefont{J.}~\bibnamefont{Ishizuka}}, \bibinfo{author}{\bibfnamefont{T.}~\bibnamefont{Arakawa}}, \bibinfo{author}{\bibfnamefont{Y.}~\bibnamefont{Shiota}}, \bibinfo{author}{\bibfnamefont{T.}~\bibnamefont{Moriyama}}, \bibinfo{author}{\bibfnamefont{Y.}~\bibnamefont{Yanase}}, \bibnamefont{and} \bibinfo{author}{\bibfnamefont{T.}~\bibnamefont{Ono}}, \href{https://www.nature.com/articles/s41586-020-2590-4}{\bibinfo{title}{Observation of superconducting diode effect}}, \bibinfo{journal}{Nature} \textbf{\bibinfo{volume}{584}}, \bibinfo{pages}{373} (\bibinfo{year}{2020}).

\bibitem[{\citenamefont{He et~al.}(2022)\citenamefont{He, Tanaka, and Nagaosa}}]{he_phenomenological_2022}
\bibinfo{author}{\bibfnamefont{J.~J.} \bibnamefont{He}}, \bibinfo{author}{\bibfnamefont{Y.}~\bibnamefont{Tanaka}}, \bibnamefont{and} \bibinfo{author}{\bibfnamefont{N.}~\bibnamefont{Nagaosa}}, \href{https://iopscience.iop.org/article/10.1088/1367-2630/ac6766/meta}{\bibinfo{title}{A phenomenological theory of superconductor diodes}}, \bibinfo{journal}{New Journal of Physics} \textbf{\bibinfo{volume}{24}}, \bibinfo{pages}{053014} (\bibinfo{year}{2022}).

\bibitem[{\citenamefont{Yuan and Fu}(2022)}]{yuan_supercurrent_2022}
\bibinfo{author}{\bibfnamefont{N.~F.} \bibnamefont{Yuan}} \bibnamefont{and} \bibinfo{author}{\bibfnamefont{L.}~\bibnamefont{Fu}}, \href{https://www.pnas.org/doi/10.1073/pnas.2119548119}{\bibinfo{title}{Supercurrent diode effect and finite-momentum superconductors}}, \bibinfo{journal}{Proceedings of the National Academy of Sciences} \textbf{\bibinfo{volume}{119}}, \bibinfo{pages}{e2119548119} (\bibinfo{year}{2022}).

\bibitem[{\citenamefont{Zhang et~al.}(2022)\citenamefont{Zhang, Gu, Li, Hu, and Jiang}}]{zhang_general_2022}
\bibinfo{author}{\bibfnamefont{Y.}~\bibnamefont{Zhang}}, \bibinfo{author}{\bibfnamefont{Y.}~\bibnamefont{Gu}}, \bibinfo{author}{\bibfnamefont{P.}~\bibnamefont{Li}}, \bibinfo{author}{\bibfnamefont{J.}~\bibnamefont{Hu}}, \bibnamefont{and} \bibinfo{author}{\bibfnamefont{K.}~\bibnamefont{Jiang}}, \href{https://link.aps.org/doi/10.1103/PhysRevX.12.041013}{\bibinfo{title}{General {Theory} of {Josephson} {Diodes}}}, \bibinfo{journal}{Physical Review X} \textbf{\bibinfo{volume}{12}}, \bibinfo{pages}{041013} (\bibinfo{year}{2022}).

\bibitem[{\citenamefont{Misaki and Nagaosa}(2021)}]{misaki_theory_2021}
\bibinfo{author}{\bibfnamefont{K.}~\bibnamefont{Misaki}} \bibnamefont{and} \bibinfo{author}{\bibfnamefont{N.}~\bibnamefont{Nagaosa}}, \href{https://link.aps.org/doi/10.1103/PhysRevB.103.245302}{\bibinfo{title}{Theory of the nonreciprocal {Josephson} effect}}, \bibinfo{journal}{Physical Review B} \textbf{\bibinfo{volume}{103}}, \bibinfo{pages}{245302} (\bibinfo{year}{2021}).

\bibitem[{\citenamefont{Davydova et~al.}(2022)\citenamefont{Davydova, Prembabu, and Fu}}]{davydova_universal_2022}
\bibinfo{author}{\bibfnamefont{M.}~\bibnamefont{Davydova}}, \bibinfo{author}{\bibfnamefont{S.}~\bibnamefont{Prembabu}}, \bibnamefont{and} \bibinfo{author}{\bibfnamefont{L.}~\bibnamefont{Fu}}, \href{https://www.science.org/doi/full/10.1126/sciadv.abo0309}{\bibinfo{title}{Universal {Josephson} diode effect}}, \bibinfo{journal}{Science advances} \textbf{\bibinfo{volume}{8}}, \bibinfo{pages}{eabo0309} (\bibinfo{year}{2022}).

\bibitem[{\citenamefont{Souto et~al.}(2024)\citenamefont{Souto, Leijnse, Schrade, Valentini, Katsaros, and Danon}}]{souto2024tuning}
\bibinfo{author}{\bibfnamefont{R.~S.} \bibnamefont{Souto}}, \bibinfo{author}{\bibfnamefont{M.}~\bibnamefont{Leijnse}}, \bibinfo{author}{\bibfnamefont{C.}~\bibnamefont{Schrade}}, \bibinfo{author}{\bibfnamefont{M.}~\bibnamefont{Valentini}}, \bibinfo{author}{\bibfnamefont{G.}~\bibnamefont{Katsaros}}, \bibnamefont{and} \bibinfo{author}{\bibfnamefont{J.}~\bibnamefont{Danon}}, \href{https://journals.aps.org/prresearch/abstract/10.1103/PhysRevResearch.6.L022002}{\bibinfo{title}{Tuning the {Josephson} diode response with an ac current}}, \bibinfo{journal}{Physical Review Research} \textbf{\bibinfo{volume}{6}}, \bibinfo{pages}{L022002} (\bibinfo{year}{2024}).

\bibitem[{\citenamefont{Zazunov et~al.}(2009)\citenamefont{Zazunov, Egger, Jonckheere, and Martin}}]{zazunov2009anomalous}
\bibinfo{author}{\bibfnamefont{A.}~\bibnamefont{Zazunov}}, \bibinfo{author}{\bibfnamefont{R.}~\bibnamefont{Egger}}, \bibinfo{author}{\bibfnamefont{T.}~\bibnamefont{Jonckheere}}, \bibnamefont{and} \bibinfo{author}{\bibfnamefont{T.}~\bibnamefont{Martin}}, \href{https://journals.aps.org/prl/abstract/10.1103/PhysRevLett.103.147004}{\bibinfo{title}{Anomalous {Josephson} current through a spin-orbit coupled quantum dot}}, \bibinfo{journal}{Physical Review Letters} \textbf{\bibinfo{volume}{103}}, \bibinfo{pages}{147004} (\bibinfo{year}{2009}).

\bibitem[{\citenamefont{Brunetti et~al.}(2013)\citenamefont{Brunetti, Zazunov, Kundu, and Egger}}]{brunetti2013anomalous}
\bibinfo{author}{\bibfnamefont{A.}~\bibnamefont{Brunetti}}, \bibinfo{author}{\bibfnamefont{A.}~\bibnamefont{Zazunov}}, \bibinfo{author}{\bibfnamefont{A.}~\bibnamefont{Kundu}}, \bibnamefont{and} \bibinfo{author}{\bibfnamefont{R.}~\bibnamefont{Egger}}, \href{https://journals.aps.org/prb/abstract/10.1103/PhysRevB.88.144515}{\bibinfo{title}{Anomalous {Josephson} current, incipient time-reversal symmetry breaking, and {Majorana} bound states in interacting multilevel dots}}, \bibinfo{journal}{Physical Review B} \textbf{\bibinfo{volume}{88}}, \bibinfo{pages}{144515} (\bibinfo{year}{2013}).

\bibitem[{\citenamefont{Pal et~al.}(2022)\citenamefont{Pal, Chakraborty, Sivakumar, Davydova, Gopi, Pandeya, Krieger, Zhang, Date, Ju et~al.}}]{pal2022josephson}
\bibinfo{author}{\bibfnamefont{B.}~\bibnamefont{Pal}}, \bibinfo{author}{\bibfnamefont{A.}~\bibnamefont{Chakraborty}}, \bibinfo{author}{\bibfnamefont{P.~K.} \bibnamefont{Sivakumar}}, \bibinfo{author}{\bibfnamefont{M.}~\bibnamefont{Davydova}}, \bibinfo{author}{\bibfnamefont{A.~K.} \bibnamefont{Gopi}}, \bibinfo{author}{\bibfnamefont{A.~K.} \bibnamefont{Pandeya}}, \bibinfo{author}{\bibfnamefont{J.~A.} \bibnamefont{Krieger}}, \bibinfo{author}{\bibfnamefont{Y.}~\bibnamefont{Zhang}}, \bibinfo{author}{\bibfnamefont{M.}~\bibnamefont{Date}}, \bibinfo{author}{\bibfnamefont{S.}~\bibnamefont{Ju}}, \bibnamefont{et~al.}, \href{https://www.nature.com/articles/s41567-022-01699-5}{\bibinfo{title}{Josephson diode effect from {Cooper} pair momentum in a topological semimetal}}, \bibinfo{journal}{Nature Physics} \textbf{\bibinfo{volume}{18}}, \bibinfo{pages}{1228} (\bibinfo{year}{2022}).

\bibitem[{\citenamefont{Baumgartner et~al.}(2022)\citenamefont{Baumgartner, Fuchs, Costa, Reinhardt, Gronin, Gardner, Lindemann, Manfra, Faria~Junior, Kochan et~al.}}]{baumgartner_supercurrent_2022}
\bibinfo{author}{\bibfnamefont{C.}~\bibnamefont{Baumgartner}}, \bibinfo{author}{\bibfnamefont{L.}~\bibnamefont{Fuchs}}, \bibinfo{author}{\bibfnamefont{A.}~\bibnamefont{Costa}}, \bibinfo{author}{\bibfnamefont{S.}~\bibnamefont{Reinhardt}}, \bibinfo{author}{\bibfnamefont{S.}~\bibnamefont{Gronin}}, \bibinfo{author}{\bibfnamefont{G.~C.} \bibnamefont{Gardner}}, \bibinfo{author}{\bibfnamefont{T.}~\bibnamefont{Lindemann}}, \bibinfo{author}{\bibfnamefont{M.~J.} \bibnamefont{Manfra}}, \bibinfo{author}{\bibfnamefont{P.~E.} \bibnamefont{Faria~Junior}}, \bibinfo{author}{\bibfnamefont{D.}~\bibnamefont{Kochan}}, \bibnamefont{et~al.}, \href{https://www.nature.com/articles/s41565-021-01009-9}{\bibinfo{title}{Supercurrent rectification and magnetochiral effects in symmetric {Josephson} junctions}}, \bibinfo{journal}{Nature Nanotechnology} \textbf{\bibinfo{volume}{17}}, \bibinfo{pages}{39} (\bibinfo{year}{2022}).

\bibitem[{\citenamefont{Legg et~al.}(2022)\citenamefont{Legg, Loss, and Klinovaja}}]{legg2022superconducting}
\bibinfo{author}{\bibfnamefont{H.~F.} \bibnamefont{Legg}}, \bibinfo{author}{\bibfnamefont{D.}~\bibnamefont{Loss}}, \bibnamefont{and} \bibinfo{author}{\bibfnamefont{J.}~\bibnamefont{Klinovaja}}, \href{https://journals.aps.org/prb/abstract/10.1103/PhysRevB.106.104501}{\bibinfo{title}{Superconducting diode effect due to magnetochiral anisotropy in topological insulators and {Rashba} nanowires}}, \bibinfo{journal}{Physical Review B} \textbf{\bibinfo{volume}{106}}, \bibinfo{pages}{104501} (\bibinfo{year}{2022}).

\bibitem[{\citenamefont{Lotfizadeh et~al.}(2024)\citenamefont{Lotfizadeh, Schiela, Pekerten, Yu, Elfeky, Strickland, Matos-Abiague, and Shabani}}]{lotfizadeh2023superconducting}
\bibinfo{author}{\bibfnamefont{N.}~\bibnamefont{Lotfizadeh}}, \bibinfo{author}{\bibfnamefont{W.~F.} \bibnamefont{Schiela}}, \bibinfo{author}{\bibfnamefont{B.}~\bibnamefont{Pekerten}}, \bibinfo{author}{\bibfnamefont{P.}~\bibnamefont{Yu}}, \bibinfo{author}{\bibfnamefont{B.~H.} \bibnamefont{Elfeky}}, \bibinfo{author}{\bibfnamefont{W.~M.} \bibnamefont{Strickland}}, \bibinfo{author}{\bibfnamefont{A.}~\bibnamefont{Matos-Abiague}}, \bibnamefont{and} \bibinfo{author}{\bibfnamefont{J.}~\bibnamefont{Shabani}}, \href{https://www.nature.com/articles/s42005-024-01618-5}{\bibinfo{title}{Superconducting diode effect sign change in epitaxial {Al}-{InAs} {Josephson} junctions}}, \bibinfo{journal}{Communications Physics} \textbf{\bibinfo{volume}{7}}, \bibinfo{pages}{120} (\bibinfo{year}{2024}).

\bibitem[{\citenamefont{Costa et~al.}(2023)\citenamefont{Costa, Fabian, and Kochan}}]{costa2023microscopic}
\bibinfo{author}{\bibfnamefont{A.}~\bibnamefont{Costa}}, \bibinfo{author}{\bibfnamefont{J.}~\bibnamefont{Fabian}}, \bibnamefont{and} \bibinfo{author}{\bibfnamefont{D.}~\bibnamefont{Kochan}}, \href{https://journals.aps.org/prb/abstract/10.1103/PhysRevB.108.054522}{\bibinfo{title}{Microscopic study of the josephson supercurrent diode effect in {Josephson} junctions based on two-dimensional electron gas}}, \bibinfo{journal}{Physical Review B} \textbf{\bibinfo{volume}{108}}, \bibinfo{pages}{054522} (\bibinfo{year}{2023}).

\bibitem[{\citenamefont{Maiani et~al.}(2023)\citenamefont{Maiani, Flensberg, Leijnse, Schrade, Vaitiek{\.e}nas, and Souto}}]{maiani2023nonsinusoidal}
\bibinfo{author}{\bibfnamefont{A.}~\bibnamefont{Maiani}}, \bibinfo{author}{\bibfnamefont{K.}~\bibnamefont{Flensberg}}, \bibinfo{author}{\bibfnamefont{M.}~\bibnamefont{Leijnse}}, \bibinfo{author}{\bibfnamefont{C.}~\bibnamefont{Schrade}}, \bibinfo{author}{\bibfnamefont{S.}~\bibnamefont{Vaitiek{\.e}nas}}, \bibnamefont{and} \bibinfo{author}{\bibfnamefont{R.~S.} \bibnamefont{Souto}}, \href{https://journals.aps.org/prb/abstract/10.1103/PhysRevB.107.245415}{\bibinfo{title}{Nonsinusoidal current-phase relations in semiconductor--superconductor--ferromagnetic insulator devices}}, \bibinfo{journal}{Physical Review B} \textbf{\bibinfo{volume}{107}}, \bibinfo{pages}{245415} (\bibinfo{year}{2023}).

\bibitem[{\citenamefont{Hess et~al.}(2023)\citenamefont{Hess, Legg, Loss, and Klinovaja}}]{hess2023josephson}
\bibinfo{author}{\bibfnamefont{R.}~\bibnamefont{Hess}}, \bibinfo{author}{\bibfnamefont{H.~F.} \bibnamefont{Legg}}, \bibinfo{author}{\bibfnamefont{D.}~\bibnamefont{Loss}}, \bibnamefont{and} \bibinfo{author}{\bibfnamefont{J.}~\bibnamefont{Klinovaja}}, \href{https://journals.aps.org/prb/abstract/10.1103/PhysRevB.108.174516}{\bibinfo{title}{Josephson transistor from the superconducting diode effect in domain wall and skyrmion magnetic racetracks}}, \bibinfo{journal}{Physical Review B} \textbf{\bibinfo{volume}{108}}, \bibinfo{pages}{174516} (\bibinfo{year}{2023}).

\bibitem[{\citenamefont{Hou et~al.}(2023)\citenamefont{Hou, Nichele, Chi, Lodesani, Wu, Ritter, Haxell, Davydova, Ili{\'c}, Glezakou-Elbert et~al.}}]{hou2023ubiquitous}
\bibinfo{author}{\bibfnamefont{Y.}~\bibnamefont{Hou}}, \bibinfo{author}{\bibfnamefont{F.}~\bibnamefont{Nichele}}, \bibinfo{author}{\bibfnamefont{H.}~\bibnamefont{Chi}}, \bibinfo{author}{\bibfnamefont{A.}~\bibnamefont{Lodesani}}, \bibinfo{author}{\bibfnamefont{Y.}~\bibnamefont{Wu}}, \bibinfo{author}{\bibfnamefont{M.~F.} \bibnamefont{Ritter}}, \bibinfo{author}{\bibfnamefont{D.~Z.} \bibnamefont{Haxell}}, \bibinfo{author}{\bibfnamefont{M.}~\bibnamefont{Davydova}}, \bibinfo{author}{\bibfnamefont{S.}~\bibnamefont{Ili{\'c}}}, \bibinfo{author}{\bibfnamefont{O.}~\bibnamefont{Glezakou-Elbert}}, \bibnamefont{et~al.}, \href{https://journals.aps.org/prl/abstract/10.1103/PhysRevLett.131.027001}{\bibinfo{title}{Ubiquitous superconducting diode effect in superconductor thin films}}, \bibinfo{journal}{Physical Review Letters} \textbf{\bibinfo{volume}{131}}, \bibinfo{pages}{027001} (\bibinfo{year}{2023}).

\bibitem[{\citenamefont{Kononov et~al.}(2020)\citenamefont{Kononov, Abulizi, Qu, Yan, Mandrus, Watanabe, Taniguchi, and Sch\"onenberger}}]{kononov2020one}
\bibinfo{author}{\bibfnamefont{A.}~\bibnamefont{Kononov}}, \bibinfo{author}{\bibfnamefont{G.}~\bibnamefont{Abulizi}}, \bibinfo{author}{\bibfnamefont{K.}~\bibnamefont{Qu}}, \bibinfo{author}{\bibfnamefont{J.}~\bibnamefont{Yan}}, \bibinfo{author}{\bibfnamefont{D.}~\bibnamefont{Mandrus}}, \bibinfo{author}{\bibfnamefont{K.}~\bibnamefont{Watanabe}}, \bibinfo{author}{\bibfnamefont{T.}~\bibnamefont{Taniguchi}}, \bibnamefont{and} \bibinfo{author}{\bibfnamefont{C.}~\bibnamefont{Sch\"onenberger}}, \href{https://pubs.acs.org/doi/10.1021/acs.nanolett.0c00658}{\bibinfo{title}{One-dimensional edge transport in few-layer $\text{WTe}_2$}}, \bibinfo{journal}{Nano letters} \textbf{\bibinfo{volume}{20}}, \bibinfo{pages}{4228} (\bibinfo{year}{2020}).

\bibitem[{\citenamefont{Souto et~al.}(2022)\citenamefont{Souto, Leijnse, and Schrade}}]{souto2022josephson}
\bibinfo{author}{\bibfnamefont{R.~S.} \bibnamefont{Souto}}, \bibinfo{author}{\bibfnamefont{M.}~\bibnamefont{Leijnse}}, \bibnamefont{and} \bibinfo{author}{\bibfnamefont{C.}~\bibnamefont{Schrade}}, \href{https://journals.aps.org/prl/abstract/10.1103/PhysRevLett.129.267702}{\bibinfo{title}{Josephson {Diode} {Effect} in {Supercurrent} {Interferometers}}}, \bibinfo{journal}{Physical Review Letters} \textbf{\bibinfo{volume}{129}}, \bibinfo{pages}{267702} (\bibinfo{year}{2022}).

\bibitem[{\citenamefont{Gupta et~al.}(2023)\citenamefont{Gupta, Graziano, Pendharkar, Dong, Dempsey, Palmstr{\o}m, and Pribiag}}]{gupta2023gate}
\bibinfo{author}{\bibfnamefont{M.}~\bibnamefont{Gupta}}, \bibinfo{author}{\bibfnamefont{G.~V.} \bibnamefont{Graziano}}, \bibinfo{author}{\bibfnamefont{M.}~\bibnamefont{Pendharkar}}, \bibinfo{author}{\bibfnamefont{J.~T.} \bibnamefont{Dong}}, \bibinfo{author}{\bibfnamefont{C.~P.} \bibnamefont{Dempsey}}, \bibinfo{author}{\bibfnamefont{C.}~\bibnamefont{Palmstr{\o}m}}, \bibnamefont{and} \bibinfo{author}{\bibfnamefont{V.~S.} \bibnamefont{Pribiag}}, \href{https://www.nature.com/articles/s41467-023-38856-0}{\bibinfo{title}{Gate-tunable superconducting diode effect in a three-terminal {Josephson} device}}, \bibinfo{journal}{Nature Communications} \textbf{\bibinfo{volume}{14}}, \bibinfo{pages}{3078} (\bibinfo{year}{2023}).

\bibitem[{\citenamefont{Ciaccia et~al.}(2023)\citenamefont{Ciaccia, Haller, Drachmann, Lindemann, Manfra, Schrade, and Sch\"onenberger}}]{ciaccia2023gate}
\bibinfo{author}{\bibfnamefont{C.}~\bibnamefont{Ciaccia}}, \bibinfo{author}{\bibfnamefont{R.}~\bibnamefont{Haller}}, \bibinfo{author}{\bibfnamefont{A.~C.~C.} \bibnamefont{Drachmann}}, \bibinfo{author}{\bibfnamefont{T.}~\bibnamefont{Lindemann}}, \bibinfo{author}{\bibfnamefont{M.~J.} \bibnamefont{Manfra}}, \bibinfo{author}{\bibfnamefont{C.}~\bibnamefont{Schrade}}, \bibnamefont{and} \bibinfo{author}{\bibfnamefont{C.}~\bibnamefont{Sch\"onenberger}}, \href{https://link.aps.org/doi/10.1103/PhysRevResearch.5.033131}{\bibinfo{title}{Gate-tunable {Josephson} diode in proximitized {InAs} supercurrent interferometers}}, \bibinfo{journal}{Phys. Rev. Res.} \textbf{\bibinfo{volume}{5}}, \bibinfo{pages}{033131} (\bibinfo{year}{2023}).

\bibitem[{\citenamefont{Valentini et~al.}(2024)\citenamefont{Valentini, Sagi, Baghumyan, de~Gijsel, Jung, Calcaterra, Ballabio, Aguilera~Servin, Aggarwal, Janik et~al.}}]{valentini2023radio}
\bibinfo{author}{\bibfnamefont{M.}~\bibnamefont{Valentini}}, \bibinfo{author}{\bibfnamefont{O.}~\bibnamefont{Sagi}}, \bibinfo{author}{\bibfnamefont{L.}~\bibnamefont{Baghumyan}}, \bibinfo{author}{\bibfnamefont{T.}~\bibnamefont{de~Gijsel}}, \bibinfo{author}{\bibfnamefont{J.}~\bibnamefont{Jung}}, \bibinfo{author}{\bibfnamefont{S.}~\bibnamefont{Calcaterra}}, \bibinfo{author}{\bibfnamefont{A.}~\bibnamefont{Ballabio}}, \bibinfo{author}{\bibfnamefont{J.}~\bibnamefont{Aguilera~Servin}}, \bibinfo{author}{\bibfnamefont{K.}~\bibnamefont{Aggarwal}}, \bibinfo{author}{\bibfnamefont{M.}~\bibnamefont{Janik}}, \bibnamefont{et~al.}, \href{https://www.nature.com/articles/s41467-023-44114-0}{\bibinfo{title}{Parity-conserving {Cooper}-pair transport and ideal superconducting diode in planar germanium}}, \bibinfo{journal}{Nature Communications} \textbf{\bibinfo{volume}{15}}, \bibinfo{pages}{169} (\bibinfo{year}{2024}).

\bibitem[{\citenamefont{Greco et~al.}(2023)\citenamefont{Greco, Pichard, and Giazotto}}]{greco2023josephson}
\bibinfo{author}{\bibfnamefont{A.}~\bibnamefont{Greco}}, \bibinfo{author}{\bibfnamefont{Q.}~\bibnamefont{Pichard}}, \bibnamefont{and} \bibinfo{author}{\bibfnamefont{F.}~\bibnamefont{Giazotto}}, \href{https://pubs.aip.org/aip/apl/article/123/9/092601/2908308/Josephson-diode-effect-in-monolithic-dc-SQUIDs}{\bibinfo{title}{Josephson diode effect in monolithic dc-{SQUIDs} based on {3D} {Dayem} nanobridges}}, \bibinfo{journal}{Applied Physics Letters} \textbf{\bibinfo{volume}{123}} (\bibinfo{year}{2023}).

\bibitem[{\citenamefont{Legg et~al.}(2023)\citenamefont{Legg, Laubscher, Loss, and Klinovaja}}]{legg2023parity}
\bibinfo{author}{\bibfnamefont{H.~F.} \bibnamefont{Legg}}, \bibinfo{author}{\bibfnamefont{K.}~\bibnamefont{Laubscher}}, \bibinfo{author}{\bibfnamefont{D.}~\bibnamefont{Loss}}, \bibnamefont{and} \bibinfo{author}{\bibfnamefont{J.}~\bibnamefont{Klinovaja}}, \href{https://journals.aps.org/prb/abstract/10.1103/PhysRevB.108.214520}{\bibinfo{title}{Parity-protected superconducting diode effect in topological {Josephson} junctions}}, \bibinfo{journal}{Physical Review B} \textbf{\bibinfo{volume}{108}}, \bibinfo{pages}{214520} (\bibinfo{year}{2023}).

\bibitem[{\citenamefont{Cuozzo et~al.}(2024)\citenamefont{Cuozzo, Pan, Shabani, and Rossi}}]{cuozzo2023microwave}
\bibinfo{author}{\bibfnamefont{J.~J.} \bibnamefont{Cuozzo}}, \bibinfo{author}{\bibfnamefont{W.}~\bibnamefont{Pan}}, \bibinfo{author}{\bibfnamefont{J.}~\bibnamefont{Shabani}}, \bibnamefont{and} \bibinfo{author}{\bibfnamefont{E.}~\bibnamefont{Rossi}}, \href{https://journals.aps.org/prresearch/abstract/10.1103/PhysRevResearch.6.023011}{\bibinfo{title}{Microwave-tunable diode effect in asymmetric {SQUIDs} with topological {Josephson} junctions}}, \bibinfo{journal}{Physical Review Research} \textbf{\bibinfo{volume}{6}}, \bibinfo{pages}{023011} (\bibinfo{year}{2024}).

\bibitem[{\citenamefont{Schrade and Fatemi}(2024)}]{PhysRevApplied.21.064029}
\bibinfo{author}{\bibfnamefont{C.}~\bibnamefont{Schrade}} \bibnamefont{and} \bibinfo{author}{\bibfnamefont{V.}~\bibnamefont{Fatemi}}, \href{https://link.aps.org/doi/10.1103/PhysRevApplied.21.064029}{\bibinfo{title}{Dissipationless nonlinearity in quantum material josephson diodes}}, \bibinfo{journal}{Phys. Rev. Appl.} \textbf{\bibinfo{volume}{21}}, \bibinfo{pages}{064029} (\bibinfo{year}{2024}).

\bibitem[{\citenamefont{Lin et~al.}(2022)\citenamefont{Lin, Siriviboon, Scammell, Liu, Rhodes, Watanabe, Taniguchi, Hone, Scheurer, and Li}}]{ZeroFieldDiode}
\bibinfo{author}{\bibfnamefont{J.-X.} \bibnamefont{Lin}}, \bibinfo{author}{\bibfnamefont{P.}~\bibnamefont{Siriviboon}}, \bibinfo{author}{\bibfnamefont{H.~D.} \bibnamefont{Scammell}}, \bibinfo{author}{\bibfnamefont{S.}~\bibnamefont{Liu}}, \bibinfo{author}{\bibfnamefont{D.}~\bibnamefont{Rhodes}}, \bibinfo{author}{\bibfnamefont{K.}~\bibnamefont{Watanabe}}, \bibinfo{author}{\bibfnamefont{T.}~\bibnamefont{Taniguchi}}, \bibinfo{author}{\bibfnamefont{J.}~\bibnamefont{Hone}}, \bibinfo{author}{\bibfnamefont{M.~S.} \bibnamefont{Scheurer}}, \bibnamefont{and} \bibinfo{author}{\bibfnamefont{J.~I.~A.} \bibnamefont{Li}}, \href{https://doi.org/10.1038/s41567-022-01700-1}{\bibinfo{title}{Zero-field superconducting diode effect in small-twist-angle trilayer graphene}}, \bibinfo{journal}{Nature Physics} \textbf{\bibinfo{volume}{18}}, \bibinfo{pages}{1221} (\bibinfo{year}{2022}).

\bibitem[{\citenamefont{Scammell et~al.}(2022)\citenamefont{Scammell, Li, and Scheurer}}]{scammell_theory_2022}
\bibinfo{author}{\bibfnamefont{H.~D.} \bibnamefont{Scammell}}, \bibinfo{author}{\bibfnamefont{J.}~\bibnamefont{Li}}, \bibnamefont{and} \bibinfo{author}{\bibfnamefont{M.~S.} \bibnamefont{Scheurer}}, \href{https://iopscience.iop.org/article/10.1088/2053-1583/ac5b16}{\bibinfo{title}{Theory of zero-field superconducting diode effect in twisted trilayer graphene}}, \bibinfo{journal}{2D Materials} \textbf{\bibinfo{volume}{9}}, \bibinfo{pages}{025027} (\bibinfo{year}{2022}).

\bibitem[{\citenamefont{Wu et~al.}(2022)\citenamefont{Wu, Wang, Xu, Sivakumar, Pasco, Filippozzi, Parkin, Zeng, McQueen, and Ali}}]{wu_field-free_2022}
\bibinfo{author}{\bibfnamefont{H.}~\bibnamefont{Wu}}, \bibinfo{author}{\bibfnamefont{Y.}~\bibnamefont{Wang}}, \bibinfo{author}{\bibfnamefont{Y.}~\bibnamefont{Xu}}, \bibinfo{author}{\bibfnamefont{P.~K.} \bibnamefont{Sivakumar}}, \bibinfo{author}{\bibfnamefont{C.}~\bibnamefont{Pasco}}, \bibinfo{author}{\bibfnamefont{U.}~\bibnamefont{Filippozzi}}, \bibinfo{author}{\bibfnamefont{S.~S.~P.} \bibnamefont{Parkin}}, \bibinfo{author}{\bibfnamefont{Y.-J.} \bibnamefont{Zeng}}, \bibinfo{author}{\bibfnamefont{T.}~\bibnamefont{McQueen}}, \bibnamefont{and} \bibinfo{author}{\bibfnamefont{M.~N.} \bibnamefont{Ali}}, \href{https://www.nature.com/articles/s41586-022-04504-8}{\bibinfo{title}{The field-free {Josephson} diode in a van der {Waals} heterostructure}}, \bibinfo{journal}{Nature} \textbf{\bibinfo{volume}{604}}, \bibinfo{pages}{653} (\bibinfo{year}{2022}).

\bibitem[{\citenamefont{Zhang et~al.}(2024{\natexlab{b}})\citenamefont{Zhang, Lin, Chichinadze, Wang, Watanabe, Taniguchi, Fu, and Li}}]{zhang2024angle}
\bibinfo{author}{\bibfnamefont{N.~J.} \bibnamefont{Zhang}}, \bibinfo{author}{\bibfnamefont{J.-X.} \bibnamefont{Lin}}, \bibinfo{author}{\bibfnamefont{D.~V.} \bibnamefont{Chichinadze}}, \bibinfo{author}{\bibfnamefont{Y.}~\bibnamefont{Wang}}, \bibinfo{author}{\bibfnamefont{K.}~\bibnamefont{Watanabe}}, \bibinfo{author}{\bibfnamefont{T.}~\bibnamefont{Taniguchi}}, \bibinfo{author}{\bibfnamefont{L.}~\bibnamefont{Fu}}, \bibnamefont{and} \bibinfo{author}{\bibfnamefont{J.}~\bibnamefont{Li}}, \href{https://www.nature.com/articles/s41563-024-01809-z}{\bibinfo{title}{Angle-resolved transport non-reciprocity and spontaneous symmetry breaking in twisted trilayer graphene}}, \bibinfo{journal}{Nature Materials} \textbf{\bibinfo{volume}{23}}, \bibinfo{pages}{356} (\bibinfo{year}{2024}{\natexlab{b}}).

\bibitem[{\citenamefont{Banerjee and Scheurer}(2024)}]{PhysRevLett.132.046003}
\bibinfo{author}{\bibfnamefont{S.}~\bibnamefont{Banerjee}} \bibnamefont{and} \bibinfo{author}{\bibfnamefont{M.~S.} \bibnamefont{Scheurer}}, \href{https://link.aps.org/doi/10.1103/PhysRevLett.132.046003}{\bibinfo{title}{Enhanced superconducting diode effect due to coexisting phases}}, \bibinfo{journal}{Phys. Rev. Lett.} \textbf{\bibinfo{volume}{132}}, \bibinfo{pages}{046003} (\bibinfo{year}{2024}).

\bibitem[{\citenamefont{D{\'\i}ez-M{\'e}rida et~al.}(2023)\citenamefont{D{\'\i}ez-M{\'e}rida, D{\'\i}ez-Carl{\'o}n, Yang, Xie, Gao, Senior, Watanabe, Taniguchi, Lu, Higginbotham et~al.}}]{diez2023symmetry}
\bibinfo{author}{\bibfnamefont{J.}~\bibnamefont{D{\'\i}ez-M{\'e}rida}}, \bibinfo{author}{\bibfnamefont{A.}~\bibnamefont{D{\'\i}ez-Carl{\'o}n}}, \bibinfo{author}{\bibfnamefont{S.}~\bibnamefont{Yang}}, \bibinfo{author}{\bibfnamefont{Y.-M.} \bibnamefont{Xie}}, \bibinfo{author}{\bibfnamefont{X.-J.} \bibnamefont{Gao}}, \bibinfo{author}{\bibfnamefont{J.}~\bibnamefont{Senior}}, \bibinfo{author}{\bibfnamefont{K.}~\bibnamefont{Watanabe}}, \bibinfo{author}{\bibfnamefont{T.}~\bibnamefont{Taniguchi}}, \bibinfo{author}{\bibfnamefont{X.}~\bibnamefont{Lu}}, \bibinfo{author}{\bibfnamefont{A.~P.} \bibnamefont{Higginbotham}}, \bibnamefont{et~al.}, \href{https://www.nature.com/articles/s41467-023-38005-7}{\bibinfo{title}{Symmetry-broken {Josephson} junctions and superconducting diodes in magic-angle twisted bilayer graphene}}, \bibinfo{journal}{Nature Communications} \textbf{\bibinfo{volume}{14}}, \bibinfo{pages}{2396} (\bibinfo{year}{2023}).

\bibitem[{\citenamefont{Hu et~al.}(2023)\citenamefont{Hu, Sun, Xie, and Law}}]{hu2023josephson}
\bibinfo{author}{\bibfnamefont{J.-X.} \bibnamefont{Hu}}, \bibinfo{author}{\bibfnamefont{Z.-T.} \bibnamefont{Sun}}, \bibinfo{author}{\bibfnamefont{Y.-M.} \bibnamefont{Xie}}, \bibnamefont{and} \bibinfo{author}{\bibfnamefont{K.~T.} \bibnamefont{Law}}, \href{https://journals.aps.org/prl/abstract/10.1103/PhysRevLett.130.266003}{\bibinfo{title}{Josephson {Diode} {Effect} {Induced} by {Valley} {Polarization} in {Twisted} {Bilayer} {Graphene}}}, \bibinfo{journal}{Physical Review Letters} \textbf{\bibinfo{volume}{130}}, \bibinfo{pages}{266003} (\bibinfo{year}{2023}).

\bibitem[{\citenamefont{Lüscher and Fischer}(2025)}]{luescher2025}
\bibinfo{author}{\bibfnamefont{B.~E.} \bibnamefont{Lüscher}} \bibnamefont{and} \bibinfo{author}{\bibfnamefont{M.~H.} \bibnamefont{Fischer}}, \href{https://arxiv.org/abs/2506.16508}{\bibinfo{title}{Superconductivity in a chern band: effect of time-reversal-symmetry breaking on superconductivity}} (\bibinfo{year}{2025}), \eprint{2506.16508}.

\bibitem[{\citenamefont{Han et~al.}(2024)\citenamefont{Han, Lu, Hadjri, Shi, Wu, Xu, Yao, Cotten, Sedeh, Weldeyesus et~al.}}]{han2024}
\bibinfo{author}{\bibfnamefont{T.}~\bibnamefont{Han}}, \bibinfo{author}{\bibfnamefont{Z.}~\bibnamefont{Lu}}, \bibinfo{author}{\bibfnamefont{Z.}~\bibnamefont{Hadjri}}, \bibinfo{author}{\bibfnamefont{L.}~\bibnamefont{Shi}}, \bibinfo{author}{\bibfnamefont{Z.}~\bibnamefont{Wu}}, \bibinfo{author}{\bibfnamefont{W.}~\bibnamefont{Xu}}, \bibinfo{author}{\bibfnamefont{Y.}~\bibnamefont{Yao}}, \bibinfo{author}{\bibfnamefont{A.~A.} \bibnamefont{Cotten}}, \bibinfo{author}{\bibfnamefont{O.~S.} \bibnamefont{Sedeh}}, \bibinfo{author}{\bibfnamefont{H.}~\bibnamefont{Weldeyesus}}, \bibnamefont{et~al.}, \href{https://arxiv.org/abs/2408.15233}{\bibinfo{title}{Signatures of chiral superconductivity in rhombohedral graphene}}, \bibinfo{journal}{arXiv preprint arXiv:2408.15233}  (\bibinfo{year}{2024}).

\bibitem[{\citenamefont{Geier et~al.}(2024)\citenamefont{Geier, Davydova, and Fu}}]{geier2024chiral}
\bibinfo{author}{\bibfnamefont{M.}~\bibnamefont{Geier}}, \bibinfo{author}{\bibfnamefont{M.}~\bibnamefont{Davydova}}, \bibnamefont{and} \bibinfo{author}{\bibfnamefont{L.}~\bibnamefont{Fu}}, \href{https://arxiv.org/abs/2409.13829}{\bibinfo{title}{Chiral and topological superconductivity in isospin polarized multilayer graphene}}, \bibinfo{journal}{arXiv preprint arXiv:2409.13829}  (\bibinfo{year}{2024}).

\bibitem[{\citenamefont{Chou et~al.}(2024)\citenamefont{Chou, Zhu, and Sarma}}]{chou2024intravalley}
\bibinfo{author}{\bibfnamefont{Y.-Z.} \bibnamefont{Chou}}, \bibinfo{author}{\bibfnamefont{J.}~\bibnamefont{Zhu}}, \bibnamefont{and} \bibinfo{author}{\bibfnamefont{S.~D.} \bibnamefont{Sarma}}, \href{https://arxiv.org/abs/2409.06701}{\bibinfo{title}{Intravalley spin-polarized superconductivity in rhombohedral tetralayer graphene}}, \bibinfo{journal}{arXiv preprint arXiv:2409.06701}  (\bibinfo{year}{2024}).

\bibitem[{\citenamefont{Yang and Zhang}(2024)}]{yang2024topological}
\bibinfo{author}{\bibfnamefont{H.}~\bibnamefont{Yang}} \bibnamefont{and} \bibinfo{author}{\bibfnamefont{Y.-H.} \bibnamefont{Zhang}}, \href{https://arxiv.org/abs/2411.02503}{\bibinfo{title}{Topological incommensurate fulde-ferrell-larkin-ovchinnikov superconductor and bogoliubov fermi surface in rhombohedral tetra-layer graphene}}, \bibinfo{journal}{arXiv preprint arXiv:2411.02503}  (\bibinfo{year}{2024}).

\bibitem[{\citenamefont{Qin and Wu}(2024)}]{qin2024chiral}
\bibinfo{author}{\bibfnamefont{Q.}~\bibnamefont{Qin}} \bibnamefont{and} \bibinfo{author}{\bibfnamefont{C.}~\bibnamefont{Wu}}, \href{https://arxiv.org/abs/2412.07145}{\bibinfo{title}{Chiral finite-momentum superconductivity in the tetralayer graphene}}, \bibinfo{journal}{arXiv preprint arXiv:2412.07145}  (\bibinfo{year}{2024}).

\bibitem[{\citenamefont{Parra-Martinez et~al.}(2025)\citenamefont{Parra-Martinez, Jimeno-Pozo, Phong, Sainz-Cruz, Kaplan, Emanuel, Oreg, Pantale{\'o}n, Silva-Guill{\'e}n, and Guinea}}]{parra2025band}
\bibinfo{author}{\bibfnamefont{G.}~\bibnamefont{Parra-Martinez}}, \bibinfo{author}{\bibfnamefont{A.}~\bibnamefont{Jimeno-Pozo}}, \bibinfo{author}{\bibfnamefont{V.~T.} \bibnamefont{Phong}}, \bibinfo{author}{\bibfnamefont{H.}~\bibnamefont{Sainz-Cruz}}, \bibinfo{author}{\bibfnamefont{D.}~\bibnamefont{Kaplan}}, \bibinfo{author}{\bibfnamefont{P.}~\bibnamefont{Emanuel}}, \bibinfo{author}{\bibfnamefont{Y.}~\bibnamefont{Oreg}}, \bibinfo{author}{\bibfnamefont{P.~A.} \bibnamefont{Pantale{\'o}n}}, \bibinfo{author}{\bibfnamefont{J.~A.} \bibnamefont{Silva-Guill{\'e}n}}, \bibnamefont{and} \bibinfo{author}{\bibfnamefont{F.}~\bibnamefont{Guinea}}, \href{https://arxiv.org/abs/2502.19474}{\bibinfo{title}{Band renormalization, quarter metals, and chiral superconductivity in rhombohedral tetralayer graphene}}, \bibinfo{journal}{arXiv preprint arXiv:2502.19474}  (\bibinfo{year}{2025}).

\bibitem[{\citenamefont{Dong and Lee}(2025)}]{dong2025}
\bibinfo{author}{\bibfnamefont{Z.}~\bibnamefont{Dong}} \bibnamefont{and} \bibinfo{author}{\bibfnamefont{P.~A.} \bibnamefont{Lee}}, \href{https://arxiv.org/abs/2503.11079}{\bibinfo{title}{A controllable theory of superconductivity due to strong repulsion in a polarized band}} (\bibinfo{year}{2025}), \eprint{2503.11079}.

\bibitem[{\citenamefont{Jahin and Lin}(2024)}]{jahin2024enhanced}
\bibinfo{author}{\bibfnamefont{A.}~\bibnamefont{Jahin}} \bibnamefont{and} \bibinfo{author}{\bibfnamefont{S.-Z.} \bibnamefont{Lin}}, \href{https://arxiv.org/abs/2411.09664}{\bibinfo{title}{Enhanced kohn-luttinger topological superconductivity in bands with nontrivial geometry}}, \bibinfo{journal}{arXiv preprint arXiv:2411.09664}  (\bibinfo{year}{2024}).

\bibitem[{\citenamefont{Wang et~al.}(2024)\citenamefont{Wang, Gao, and Yang}}]{wang2024chiral}
\bibinfo{author}{\bibfnamefont{Y.-Q.} \bibnamefont{Wang}}, \bibinfo{author}{\bibfnamefont{Z.-Q.} \bibnamefont{Gao}}, \bibnamefont{and} \bibinfo{author}{\bibfnamefont{H.}~\bibnamefont{Yang}}, \href{https://arxiv.org/abs/2410.05384}{\bibinfo{title}{Chiral superconductivity from parent chern band and its non-abelian generalization}}, \bibinfo{journal}{arXiv preprint arXiv:2410.05384}  (\bibinfo{year}{2024}).

\bibitem[{\citenamefont{Yoon et~al.}(2025)\citenamefont{Yoon, Xu, Barlas, and Zhang}}]{yoon2025quarter}
\bibinfo{author}{\bibfnamefont{C.}~\bibnamefont{Yoon}}, \bibinfo{author}{\bibfnamefont{T.}~\bibnamefont{Xu}}, \bibinfo{author}{\bibfnamefont{Y.}~\bibnamefont{Barlas}}, \bibnamefont{and} \bibinfo{author}{\bibfnamefont{F.}~\bibnamefont{Zhang}}, \href{https://arxiv.org/abs/2502.17555}{\bibinfo{title}{Quarter metal superconductivity}}, \bibinfo{journal}{arXiv preprint arXiv:2502.17555}  (\bibinfo{year}{2025}).

\bibitem[{\citenamefont{May-Mann et~al.}(2025)\citenamefont{May-Mann, Helbig, and Devakul}}]{maymann2025}
\bibinfo{author}{\bibfnamefont{J.}~\bibnamefont{May-Mann}}, \bibinfo{author}{\bibfnamefont{T.}~\bibnamefont{Helbig}}, \bibnamefont{and} \bibinfo{author}{\bibfnamefont{T.}~\bibnamefont{Devakul}}, \href{https://arxiv.org/abs/2503.05697}{\bibinfo{title}{How pairing mechanism dictates topology in valley-polarized superconductors with berry curvature}} (\bibinfo{year}{2025}), \eprint{2503.05697}.

\bibitem[{\citenamefont{Christos et~al.}(2025)\citenamefont{Christos, Bonetti, and Scheurer}}]{christos2025}
\bibinfo{author}{\bibfnamefont{M.}~\bibnamefont{Christos}}, \bibinfo{author}{\bibfnamefont{P.~M.} \bibnamefont{Bonetti}}, \bibnamefont{and} \bibinfo{author}{\bibfnamefont{M.~S.} \bibnamefont{Scheurer}}, \href{https://arxiv.org/abs/2503.15471}{\bibinfo{title}{Finite-momentum pairing and superlattice superconductivity in valley-imbalanced rhombohedral graphene}} (\bibinfo{year}{2025}), \eprint{2503.15471}.

\bibitem[{\citenamefont{Sedov and Scheurer}(2025)}]{sedov2025}
\bibinfo{author}{\bibfnamefont{D.}~\bibnamefont{Sedov}} \bibnamefont{and} \bibinfo{author}{\bibfnamefont{M.~S.} \bibnamefont{Scheurer}}, \href{https://arxiv.org/abs/2503.12650}{\bibinfo{title}{Probing superconductivity with tunneling spectroscopy in rhombohedral graphene}} (\bibinfo{year}{2025}), \eprint{2503.12650}.

\bibitem[{\citenamefont{Lesser et~al.}(2025)\citenamefont{Lesser, Huang, Sethna, and Kim}}]{lesser2025}
\bibinfo{author}{\bibfnamefont{O.}~\bibnamefont{Lesser}}, \bibinfo{author}{\bibfnamefont{C.}~\bibnamefont{Huang}}, \bibinfo{author}{\bibfnamefont{J.~P.} \bibnamefont{Sethna}}, \bibnamefont{and} \bibinfo{author}{\bibfnamefont{E.-A.} \bibnamefont{Kim}}, \href{https://arxiv.org/abs/2506.08087}{\bibinfo{title}{Emblems of pair density waves: dual identity of topological defects and their transport signatures}} (\bibinfo{year}{2025}), \eprint{2506.08087}.

\end{thebibliography}
\end{document}